\documentclass[onecolumn,aps,prd,preprintnumbers,showpacs,superscriptaddress,nofootinbib,amsmath,amssymb,floats,floatfix,showkeys,notitlepage]{revtex4-1}

\usepackage{comment}
\usepackage{graphicx}
\usepackage{subfigure}
\usepackage{palatino}
\usepackage{sans}
\usepackage{hyperref}
\hypersetup{colorlinks=true,linkcolor=blue,urlcolor=blue,citecolor=blue}
\usepackage[toc,page]{appendix}
\usepackage[normalem]{ulem}
\usepackage{adjustbox}
\usepackage{latexsym}
\usepackage{amsmath}
\usepackage{amssymb}
\usepackage{amsfonts}
\usepackage{dcolumn}
\usepackage{bm}
\usepackage{tikz}
\usepackage{bigints}
\usepackage{array,tabularx,multirow,booktabs}
\usepackage[tracking=true]{microtype}
\SetTracking{}{500}
\SetTracking{encoding={*}, shape=sc}{40}
\UseRawInputEncoding 
\allowdisplaybreaks

\newcommand{\dif}{\ensuremath{\mathrm{d}}}

\newcommand{\imag}{\ensuremath{\mathrm{i}}}

\renewcommand{\Im}{\ensuremath{\mathrm{Im}}}

\begin{document} \sloppy
\title{Black hole in quantum wave dark matter}

\author{Reggie C. Pantig}
\email{reggie.pantig@dlsu.edu.ph}
\affiliation{Physics Department, De La Salle University, 2401 Taft Avenue, Manila, 1004 Philippines}
\affiliation{Physics Department, Map\'ua University, 658 Muralla St., Intramuros, Manila 1002, Philippines}

\author{Ali \"Ovg\"un}
\email{ali.ovgun@emu.edu.tr}
\homepage{https://www.aovgun.com}
\affiliation{Physics Department, Eastern Mediterranean University, Famagusta, 99628 North Cyprus via Mersin 10, Turkey}

\begin{abstract}
In this work, we explored the effect of the fuzzy dark matter (FDM) (or wave dark matter) halo on a supermassive black hole (SMBH). Such a dark matter introduces a soliton core density profile, and we treat it ideally as a spherical distribution that surrounds the SMBH located at its center. In this direction, we obtained a new metric due to the union of the black hole and dark matter spacetime geometries. We applied the solution to the two known SMBH - Sgr. A* and M87* and used the empirical data for the shadow diameter by EHT to constrain the soliton core radius $r_\text{c}$ given some values of the boson mass $m_\text{b}$. Then, we examine the behavior of the shadow radius based on such constraints and relative to a static observer. We found that different shadow sizes are perceived at regions $r_\text{obs}<r_\text{c}$ and $r_\text{obs}>r_\text{c}$, and the deviation is greater for values $m_\text{b}<10^{-22}$ eV. Concerning the shadow behavior, we have also analyzed the effect of the soliton profile on the thin-accretion disk. Soliton dark matter effects manifest through the varying luminosity near the event horizon. We also analyzed the weak deflection angle and the produced Einstein rings due to soliton effects. We found considerable deviation, better than the shadow size deviation, for the light source near the SMBH with impact parameters comparable to the soliton core. Our results suggest the possible experimental detection of soliton dark matter effects using an SMBH at the galactic centers.
\end{abstract}


\maketitle


\section{Introduction} \label{intr}
One of the greatest mysteries of astrophysics and cosmology is the true nature of dark matter. The $\Lambda$CDM model and its pioneering success in explaining the dynamics of the large-scale Universe suggests that our Universe is made up of $27\%$ dark matter, which constitutes $85\%$ of the Universe's total mass \cite{WMAP:2010sfg}. While successful on the cosmological scale, the model faces some significant problems at the galactic scale. These notable discrepancies are: $(1)$ cusp-core problem \cite{Moore:1994yx,Flores:1994gz}, $(2)$ missing satellite problem \cite{Moore:1999nt,Klypin:1999uc}, and $(3)$ too-big-to-fail problem \cite{Boylan-Kolchin:2011qkt}.

Despite the CDM model's success and the mentioned anomalies above, one that also remained elusive is the Earth-based detection of dark matter particles associated with the CDM model which is the WIMPs (Weakly Interactive Massive Particles). Some work are promising that reported positive results \cite{DAMA:2008jlt,Bernabei:2013xsa,Bernabei:2018jrt}, but later on debunked by other testing laboratories \cite{CRESST:2015txj,PICO:2017tgi,LUX:2017ree} which found null results. Other proposed alternatives such as studying the Earth's crust years of data that may leave imprints of dark matter \cite{Baum:2018tfw}. At this time of writing, even the most sensitive dark matter detector reported that no dark matter particles are detected \cite{LZ:2022ufs}.

These aforementioned events demand the search for some new dark matter models. Useful review articles are found in Refs. \cite{Urena-Lopez:2019kud,Arbey:2021gdg,Arun:2017uaw,Kribs:2016cew}. In this paper, our interest is about the quantum wave dark matter, which is also known as the \textit{fuzzy} dark matter denoted as $\Psi$DM \cite{Schive:2014dra}. For an in-depth review, see Ref. \cite{Hui:2021tkt}. Moreover, Cardoso et al. studied the evolution of a fuzzy dark matter soliton as it is accreted by a central supermassive black hole, identifying the different stages of accretion and associated timescales \cite{Cardoso:2022nzc}. The $\Psi$DM formalism merely reconciles the cusp-core problem due to the added quantum stress repulsion at the small scale distance. On the cosmological scale, it mimics the expected behavior from the CDM model. Using spheroidal dwarf galaxies which are dark matter dominated, cosmological simulation revealed that needed solitonic boson mass of $m_\text{b} = 8.1^{+1.6}_{-1.7}\times 10^{-23}$ eV \cite{Schive:2014dra}, and more massive boson mass are expected to massive galaxies such as the Milky Way and M87 to explain spheroidal formation in its early life - a phenomenon that CDM model struggles to explain.

Since the $\Psi$DM shines on a smaller scale, it is only natural to question its effect on the black hole geometry. An answer would be important not only for galaxies that home an SMBH at their center but also to spheroidal dwarf galaxies that might have a black hole at their center waiting to be discovered. There had been several related studies in this direction. Contrary to what this paper wants to explore, Ref. \cite{Davies:2019wgi} considers the effect \textit{of} an SMBH \textit{on} the fuzzy dark matter. Recently, the effect of the Dehnen profile on certain black holes residing in a dwarf galaxy was considered \cite{Pantig:2022whj}. The effect of other dark matter profiles such as the CDM, SFDM, URC, and superfluid dark matter on the black hole geometry was also considered in Refs. \cite{Hou:2018bar,Jusufi2019,Jusufi:2020cpn}. Even more complicated models of dark matter profiles are also studied \cite{Xu:2020jpv}, as well as those having dark matter spikes \cite{Nampalliwar:2021tyz,Xu:2021dkv}. What is common in these studies is that dark matter effects due to the mentioned profiles cause an almost negligible deviation to the known shadow radius of $R_\text{sh} = 3\sqrt{3}m$. Dark matter effect on the weak deflection angle, however, causes more deviation but it demands more sensitive instruments than we currently have \cite{Pantig:2022toh,Pantig:2022whj}. Konoplya \cite{Konoplya:2019sns} also considered a dark matter toy model and examined the effect of its effective mass on the black hole shadow. Exploring the effect of such a toy model to the weak deflection angle, and spherical accretion were also studied by various authors \cite{Pantig:2020odu,Pantig:2020uhp,Pantig:2021zqe,Vagnozzi:2022moj,Ovgun:2018tua,Jusufi:2017lsl,Javed:2019kon,Javed:2019rrg,Kumaran:2019qqp,Ovgun:2020gjz,Ovgun:2020yuv,Jusufi:2017vew,Chen:2022nbb,Dymnikova2019,Uniyal:2022vdu,Kuang:2022xjp,Meng:2022kjs,Tang:2022hsu,Kuang:2022ojj,Wei2019,Xu2018a,Hou:2018avu,Bambi2019,Tsukamoto:2017fxq,Kumar:2020hgm,Kumar2019,Wang2017,Wang2018,Amarilla2018,He2020,Tsupko_2020,Hioki2009,Li2020,Ling:2021vgk,Belhaj:2020okh,Cunha:2018acu,Gralla:2019xty,Perlick:2015vta,Nedkova:2013msa,Li:2013jra,Khodadi:2021gbc,Khodadi:2022pqh,Cunha:2016wzk,Shaikh:2019fpu,Allahyari:2019jqz,Yumoto:2012kz,Cunha:2016bpi,Moffat:2015kva,Cunha:2016bjh,Zakharov:2014lqa,Hennigar:2018hza,Chakhchi:2022fl,Saurabh:2020zqg}.

With the mentioned studies above, we aim to consider the soliton dark matter profile's effect on the black hole geometry and find out whether or not it will give some notable difference to the Schwarzschild case using the known data for Sgr. A* and M87* black holes \cite{EventHorizonTelescope:2019dse,EventHorizonTelescope:2022xnr}. First, we aim to find some constraint to the value of the solitonic core radius $r_\text{c}$ given some values of the boson mass $m_\text{b}$. Then, we will explore whether the obtained parameters will give a noticeable deviation to the shadow radius relative to some static observer located at some radial distance $r_\text{o}$. In other words, we will use the black hole shadow to gain some insights into the imprints of the fuzzy dark matter on the black hole geometry. We will use mainly the formalism by Xu et. al \cite{Xu:2018wow} in obtaining the fused dark matter and black hole geometries, then the calculation of the black hole shadow through the works of \cite{Perlick:2015vta,Perlick:2021aok} which were based on the seminal papers \cite{Synge:1966okc,Luminet:1979nyg}. Another important tool in astrophysics is the gravitational lensing \cite{Virbhadra:1998dy,Virbhadra:1999nm,Virbhadra:2002ju,Virbhadra:2007kw,Virbhadra:2008ws,Adler:2022qtb,Bozza:2001xd,Bozza:2002zj,Perlick:2003vg,Virbhadra:2022ybp,Virbhadra:2022iiy}. To calculate deflection angle in weak fields, Gibbons and Werner proposed new method using the Gauss-Bonnet theorem (GBT) on the asymptotically flat optical geometries \cite{Gibbons:2008rj}. This work opens new research area in astrophysics, and it has been applied to various different phenomena  \cite{Ovgun:2018fnk,Ovgun:2019wej,Ovgun:2018oxk,Javed:2019ynm,Werner_2012,Ishihara:2016vdc,Ishihara:2016sfv,Ono:2017pie,Li:2020dln,Li:2020wvn,Belhaj:2022vte,Belhaj:2020rdb,Liu:2022lfb}. 

The program of this paper is as follows: In Sect. \ref{sec2}, the spacetime metric for a black hole surrounded by wave dark matter will be derived, where we only considered the non-rotating case. In Sect. \ref{sec3}, we study the shadow of such a black hole by first constraining the soliton mass using the EHT data, and analyze the behavior of the shadow relative to a static observer at some radial position from the black hole. As an alternative analysis to the shadow, we also study the thin-accretion disk in subsection IIIa and considered the phenomenon of weak deflection angle using GBT in subsection IIIb, and Einstein Rings in subsection IIIc. Finally, in Sect. \ref{conc}, we summarize our paper. In this paper, we used geometrized units by setting $G = c = 1$, and the metric signature as $(-,+,+,+)$.

\section{Black hole metric in quantum wave dark matter} \label{sec2}
The soliton density profile which describes the \textit{solitonic core} is given by \cite{Schive:2014dra}
\begin{equation} \label{e1}
    \rho_\text{sol}(r) = \rho_\text{c}\left[1+\alpha\left( \frac{r}{r_\text{c}}\right)^2 \right]^{-8}
\end{equation}
where $r_\text{c}$ and $\rho_\text{c}$ are defined as the soliton core radius, and soliton core density, respectively \cite{Herrera-Martin:2017cux}. Such a density profile is studied in detail in Ref. \cite{Schive:2014hza}. The core radius is defined to be the half-density comoving radius that sets exactly the constant $\alpha = \sqrt[8]{2}-1 \sim 0.09051$. In Eq. \eqref{e1}, $\rho_\text{c}$ is defined as
\begin{equation} \label{e2}
    \rho_\text{c} = 2.4\text{x}10^{12} m_\psi^{-2}  \left(\frac{r_\text{c}}{\text{pc}}\right)^{-4}\frac{M_\odot}{\text{pc}^{3}},
\end{equation}
where we wrote
\begin{equation}
    m_\psi = \frac{m_\text{b}}{10^{-22}\text{eV}}
\end{equation}
as the boson mass parameter. The boson mass $m_\text{b}$ is a fundamental parameter that is believed to have a single value for all galaxies in the Universe \cite{Herrera-Martin:2017cux}. Note that Eq. \eqref{e1} came from numerical simulations and there are no specific analytical formulae for the soliton profile. It is then only an approximation that gives convenience to compare the soliton model relative to empirical observations. In particular, the best fit for $m_\text{b}$ and $r_\text{c}$ were undertaken using observational benchmarks from popular dwarf spheroidal such as Fornax \cite{Schive:2014dra}. 

With the density profile in Eq. \eqref{e1}, we can find the mass of the core $M_\text{c}$, and the total soliton mass $M_\text{sol}$  \cite{Nori:2020jzx}:
\begin{align}
	M_{\text{c}}(r)&=4 \pi \int_{0}^{r_\text{c}} \rho_\text{sol}\left(r^{\prime}\right) r^{\prime 2} d r^{\prime} = 0.89\pi \rho_\text{c} r_\text{c}^3, \nonumber \\
	M_{\text{sol}}(r)&=4 \pi \int_{0}^{r=\infty} \rho_\text{sol}\left(r^{\prime}\right) r^{\prime 2} d r^{\prime} = 3.7184\pi \rho_\text{c} r_\text{c}^3.
\end{align}
From these results, it can be confirmed that the relation $M_{\text{sol}} = 4M_{\text{c}}$ holds. Note that such a relationship varies for different values of the exponent in Eq. \eqref{e1}. Then, scaling symmetry reveals a simple relation between $r_\text{c}$, and the mass of the core $M_\text{c}$ through \cite{Schive:2014dra}
\begin{equation} \label{e3}
    M_\text{c} = \frac{5.5\times 10^9 M_\odot}{(m_\text{b}/10^{-23} \text{eV})^2 (r_\text{sol}/\text{kpc})}.
\end{equation}
Another important relation to consider is how $M_{\text{sol}}$ is related to the virial mass of the dark matter halo $M_\text{h}$:
\begin{equation} \label{e4}
    M_\text{sol} = a^{-1/2}M_\text{h}^{1/3},
\end{equation}
where $a$ is the cosmological scale factor, which, at the present age of the Universe, takes the value of $a = 1$. Note that $M_\text{h}$ is identical to that of the CDM paradigm. Thus, with Eq. \eqref{e4}, we can determine the total mass of the soliton if we know the virial mass of the DM halo of a particular galaxy. It is useful to know the exact relation \cite{Davies:2019wgi}:
\begin{equation}
	M_\text{sol} = 1.25\times 10^9 \left(\frac{M_{\text{h}}}{10^{12}\,{ M_{\odot }}}\right)^{1/3} \left(\frac{m_\text{b}}{10^{-22}\,{\rm eV}}\right)^{-1}\,{ M_{\odot }} ,
\end{equation}

\begin{figure}
    \centering
    \includegraphics[width=0.48\textwidth]{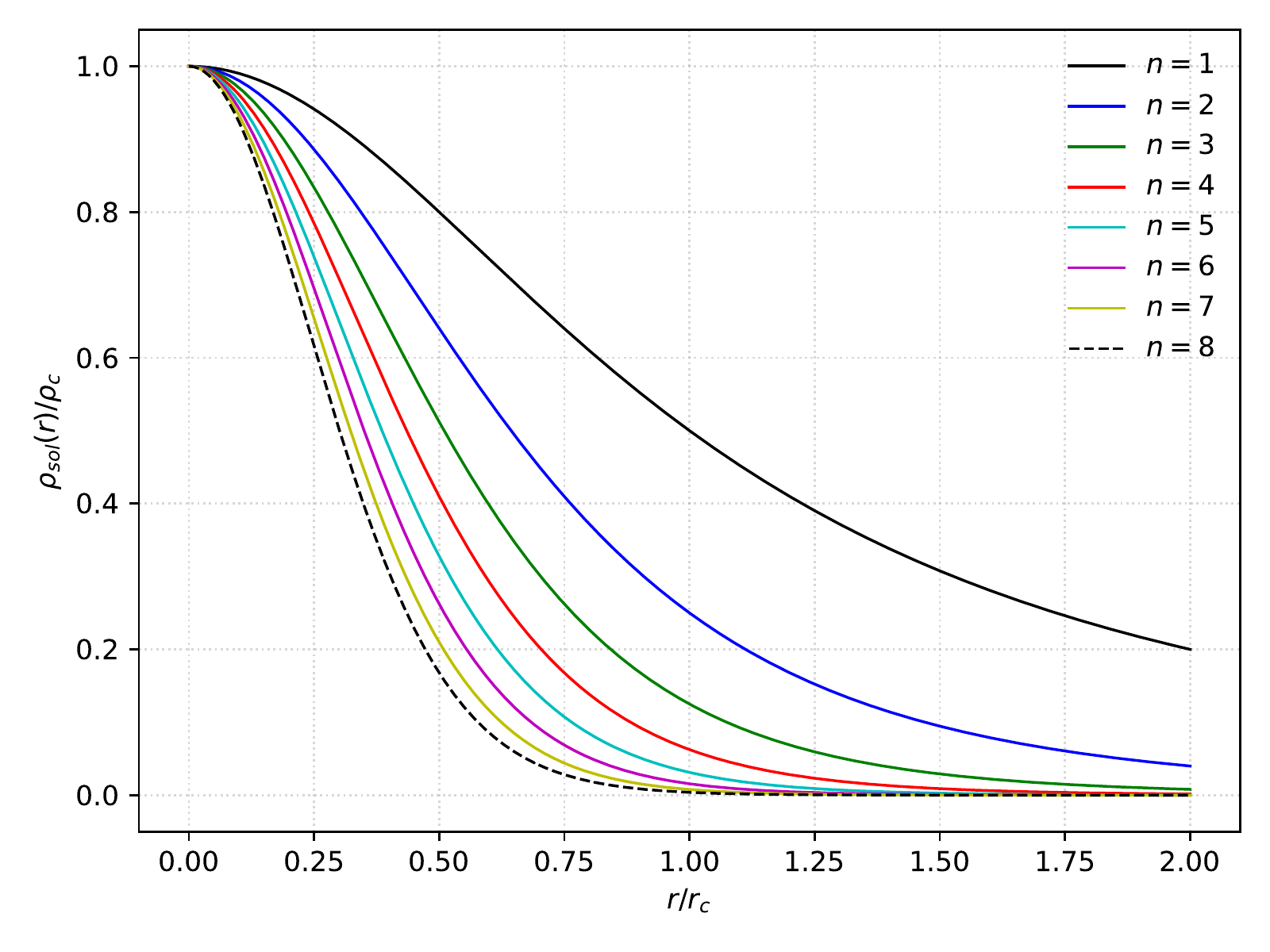}
    \caption{Plot of the density profile in Eq. \eqref{e1}.}
    \label{densprof}
\end{figure}

As all are now set, we will now derive the metric for a black hole surrounded by solitonic dark matter. To reduce the clutter in the upcoming derivation, we replace the exponent in Eq. \eqref{e1} with $n$. We can then write Eq. \eqref{e1} as
\begin{equation} \label{e6}
    \rho_\text{sol}(r) = \rho_\text{c}\left[1+\alpha\left( \frac{r}{r_\text{c}}\right)^2 \right]^{-n}.
\end{equation}
It is understood that when $n = 8$, we are considering the most important case for soliton dark matter that is frequently studied in the literature as mentioned earlier \citep{Schive:2014dra}. One may wonder, however, if we can assign a whole number integer for $n$ to explore some scenarios. Theoretically, this is possible but it may not be relevant or can be ruled out by observations. The behavior of the soliton density profile for different $n$ is shown in Fig. \ref{densprof}. For a fixed value of $r_\text{c}$ and in low values of $r$, the rate of change of the decrease in dark matter density is highest on $n = 8$. However, we can see that as $r$ further increases, $n$ from $3$ to $7$ approaches that of $n = 8$. We can see, however, that when $r = r_\text{c}$, the significant difference between these values of $n$ cannot be ignored.

Let us now formally fuse the soliton profile to the black geometry. Here, we assume that the black hole is at the center of the dark matter configuration that is spherical in shape. Here, we do not need to consider the extremes of integration and be more general. That is, we take
\begin{equation} \label{e7}
    M_{\text{sol}}(r)=4 \pi \int_{0}^{r} \rho\left(r^{\prime}\right) r^{\prime 2} d r^{\prime},
\end{equation}
which results to
\begin{equation} \label{e8}
	M_{\text{sol}}(r)=\frac{2 \pi  k}{\alpha^{\frac{3}{2}} N \lambda} \left[-2 N \left(\frac{r_\text{c}}{r}\right)^{\sigma} \alpha^{-\frac{\sigma}{2}}{}_2F_{1}\left(n,\frac{\sigma}{2};n-\frac{1}{2};-\frac{r_\text{c}^2}{r^{2} \alpha}\right) +\frac{\pi^{\frac{3}{2}} \sigma}{2} \right]
\end{equation}
after evaluation. We see the existence of a hypergeometric function. For brevity, we wrote in Eq. \eqref{e8} the following:
\begin{equation} \label{e9}
	k = \rho_\text{c}r_\text{c}^3, \qquad
	N = \Gamma(5/2-n)\Gamma(n) \cos(\pi n), \qquad
	\sigma = 2n - 3.
\end{equation}
Any test particle enveloped by the dark matter halo will then have some tangential velocity, defined by
\begin{equation} \label{e10}
	v_{\text{tg}}^{2}(r)=M_{\text{sol}}(r) / r,
\end{equation}
where Eq. \eqref{e8} must be used. Next, let us consider the line element of the dark matter halo, given as
\begin{equation} \label{e11}
    ds^{2}_{\text{h}} = -f(r) dt^{2} + g(r)^{-1} dr^{2} + r^2 d\theta ^{2} +r^2\sin^2\theta d\phi^{2}.
\end{equation}
Using such a line element, it is easy to obtain the relation between the metric function and the tangential velocity:
\begin{equation} \label{e12}
    v_{\text{tg}}(r)=r \frac{d \ln (\sqrt{f(r)})}{d r}.
\end{equation}
After using the result in Eq. \eqref{e12} and some considerable algebra, we find
\begin{equation} \label{e13}
	f(r) = \exp \Bigg\{-\frac{4 \pi  k}{\alpha^{\frac{3}{2}} \lambda N r} \left[
-h_1 N \left(\frac{r_\text{c}}{r}\right)^{\sigma} \alpha^{-\frac{\sigma}{2}}+\frac{\pi^{\frac{3}{2}} \lambda}{2}
\right]\Bigg\}
\end{equation}
where we assign $h_1$ and $\lambda$ as:
\begin{equation} \label{e14}
	h_1 = {}_2F_{1}\left(\frac{\sigma}{2},n-1;n-\frac{1}{2};-\frac{r_\text{c}^2}{r^{2} \alpha}\right), \qquad \lambda = \sigma(n-1),
\end{equation}
for the sake of brevity. Information about the dark matter profile is imprinted in $f(r)$, and one of the aims of this paper is to combine it with the black hole metric. The black hole that we are going to consider here is a static and spherically symmetric one, surrounded by the dark matter described by the density profile in Eq. \eqref{e8}. Xu et al. formalism \cite{Xu:2018wow} will aid us in such a process. The Einstein field equation, with this combination, can be modified as
\begin{equation} \label{e15}
    R^{\mu}_{\nu}=\frac{1}{2}\delta^{\mu}_{\nu}R=\kappa^2((T^{\mu}_{\nu})_{\text{DM}}+(T^{\mu}_{\nu})_{\text{Schw}}),
\end{equation}
which allows us to redefine the metric function as
\begin{equation} \label{e16}
    ds^{2} = -F(r) dt^{2} + G(r)^{-1} dr^{2} + r^2 d\theta ^{2} +r^2\sin^2\theta d\phi^{2},
\end{equation}
where we write
\begin{equation} \label{e17}
    F(r)=f(r) + F_1(r), \quad \quad G(r) = g(r)+F_2(r).
\end{equation}
As a result, Eq. \eqref{e15} gives us
\begin{align} \label{e18}
    (g(r)+F_{2}(r))\left(\frac{1}{r^{2}}+\frac{1}{r}\frac{g^{'}(r)+F^{'}_{2}(r)}{g(r)+F_{2}(r)}\right)&=g(r)\left(\frac{1}{r^{2}}+\frac{1}{r}\frac{g^{'}(r)}{g(r)}\right), \nonumber \\
    (g(r)+F_{2}(r))\left(\frac{1}{r^{2}}+\frac{1}{r}\frac{f^{'}(r)+F^{'}_{1}(r)}{f(r)+F_{1}(r)}\right)&=g(r)\left(\frac{1}{r^{2}}+\frac{1}{r}\frac{f^{'}(r)}{f(r)}\right).
\end{align}
Solving for $F_1(r)$ and $F_2(r)$ yields
\begin{align} \label{e19}
    F(r) &= \exp\left[\int \frac{g(r)}{g(r)-\frac{2m}{r}}\left(\frac{1}{r}+\frac{f^{'}(r)}{f(r)}\right)dr-\frac{1}{r} dr\right], \nonumber\\
    G(r) &=g(r)-\frac{2m}{r},
\end{align}
where $m$ is the mass of the black hole. Note that $f(r)=g(r)=1$ implies the non-existence of the dark matter halo since $k = 0$, resulting to the integral of $F(r)$ to become a constant $C_1 = 1-2m/r$. Thus, it merely reduces to the pure Schwarzschild case. The dark matter halo can be found by inspecting Eqs. \eqref{e17}-\eqref{e18}. Then, if we assume that $f(r)=g(r)$ and $F_1(r)=F_2(r)=-2m/r$, it implies that $F(r)=G(r)$, and the metric function can be simply written as 
\begin{equation}  \label{emetric}
    F(r) = \exp \Bigg\{\frac{4 \pi  k}{\alpha^{\frac{3}{2}} N \lambda r} \left[
N \left(\frac{r_\text{c}}{r}\right)^{\sigma} \alpha^{-\frac{\sigma}{2}}h_1-\frac{\lambda \pi^{\frac{3}{2}}}{2} \right]\Bigg\}-\frac{2m}{r}.
\end{equation}
In the following sections, we want to write the full metric as
\begin{equation} \label{e21}
    ds^{2} = -A(r) dt^{2} + B(r) dr^{2} + C(r) d\theta ^{2} +D(r) d\phi^{2},
\end{equation}
where $B(r)=A(r)^{-1}$, $C(r)=r^2$, and $D(r)=r^2\sin^2\theta$. With this notation, $A(r)=F(r)$, and $C(r)=D(r)$ when one wants to the spherical symmetry. Thus, it allows us to analyze the black hole geometry without loss of generality at $\theta=\pi/2$. The metric function $F(r)$ is also general since we can obtain different expressions based on the value of $n$. Aligning this study with known observations, we must choose only $n = 8$. For theoretical reasons, however, it is also interesting to consider a different value for $n$, say, $n = 7$. In the next sections, it would be useful to express $F(r)$ in a form such that $r \to \infty$. For $n = 8$, we have
\begin{equation} \label{metricc}
    F(r) = D(k,r_\text{c})-\frac{2 m}{r},
\end{equation}
where
\begin{equation} \label{ee25}
    D(k,r_\text{c}) = 1 - \frac{4\pi k}{7 \alpha r_\text{c}}.
\end{equation}

\subsection{Thermodynamics properties of the black hole}
In \cite{Abdelqader:2014vaa}, a set of curvature scalars was proposed to detect the location of the event horizon and the ergosurface as well as to define some other properties, such
as the mass and the spin of the black hole \cite{Abdelqader:2014vaa,Tavlayan:2020chf}:
\begin{equation}
\begin{array}{ll}
I_{1}=C_{\mu \nu \alpha \beta} C^{\mu \nu \alpha \beta}, & I_{2}={ }^{*} C_{\mu \nu \alpha \beta} C^{\mu \nu \alpha \beta} \\
I_{3}=\nabla_{\rho} C_{\mu \nu \alpha \beta} \nabla^{\rho} C^{\mu \nu \alpha \beta}, & I_{4}=\nabla_{\rho} C_{\mu \nu \alpha \beta} \nabla^{\rho *} C^{\mu \nu \alpha \beta} \\
I_{5}=k_{\mu} k^{\mu}, \quad I_{6}=l_{\mu} l^{\mu}, & I_{7}=k_{\mu} l^{\mu}
\end{array}
\end{equation}
where $C_{\mu \nu \alpha \beta}$ is the Weyl Tensor, ${ }^{*} C_{\mu \nu \alpha \beta}$ is its left dual and the two covectors defining the last three scalars are given as $k_{\mu}=-\nabla_{\mu} I_{1}$, and $l_{\mu}=-\nabla_{\mu} I_{2}$.

Authors of \cite{Tavlayan:2020chf} show that instead of calculating the event horizon using the largest root of $g^{r r} \equiv 0$, one can find the location of the event horizon of the Schwarzschild-like black holes, only using the $I_{3}$ because scalar vanishes on the event horizon but it is positive/negative outside/inside the event horizon. 
First, we write the induced metric in the induced coordinates $(t, r)$ as
\begin{equation}
\gamma_{i j}=\left(\begin{array}{cc}
F(r) & 0 \\
0 & \frac{1}{F(r)}.
\end{array}\right)
\end{equation}
Then we obtain the Kretschmann invariant for the induced metric
\begin{equation}
K={ }^{\Sigma} I_{1}={ }^{\Sigma} R_{i j k l}{ }^{\Sigma} R^{i j k l}= \left( {\frac {{\rm d}^{2}}{{\rm d}{r}^{2}}}F \left( r
 \right) \right) ^{2}={\frac { \left( 1024\,m{\alpha}^{3/2}+33\,{\pi}^{2}k \right) ^{2}}{
65536\,{r}^{6}{\alpha}^{3}}},
\end{equation}
to calculate the horizon detecting invariant
\begin{equation}
{ }^{\Sigma} I_{5}=\nabla_{m}{ }^{\Sigma} I_{1} \nabla^{m \Sigma} I_{1}=4\, \left( {\frac {{\rm d}^{2}}{{\rm d}{r}^{2}}}F
 \left( r \right)  \right) ^{2} \left( {\frac {{\rm d}^{3}}{{\rm d}{r}
^{3}}}F \left( r \right)  \right) ^{2}F \left( r \right)
\end{equation}
\begin{equation}
    =-\frac{18432 \left(r_c \left(m -\frac{r}{2}\right) \alpha +\frac{2 \pi  k r}{7}\right) m^{4}}{r^{15} r_c \alpha},
\end{equation}
whose largest real root, ${ }^{\Sigma} I_{5}\left(r=r_{+}\right)=0$, is exactly the event horizon. Similarly: ${ }^{\Sigma} I_{3}\left(r=r_{+}\right)=0$ one can calculate also $I_{3}$ to obtain same result:
\begin{equation}
{ }^{\Sigma} I_{3}=\nabla_{m}{ }^{\Sigma} R_{i j k l} \nabla^{m \Sigma} R^{i j k l}= \left( {\frac {{\rm d}^{3}}{{\rm d}{r}^{3}}}F \left( r
 \right)  \right) ^{2}F \left( r \right) 
\end{equation}
\begin{equation}
{ }^{\Sigma} I_{3}=-\frac{288 m^{2} \left(r_c \left(m -\frac{r}{2}\right) \alpha +\frac{2 \pi  k r}{7}\right)}{r^{9} r_c \alpha}.
\end{equation}
The event horizon is located at:
\begin{equation} \label{hor}
  r_h=-\frac{14 m r_c \alpha}{4 \pi  k -7 r_c \alpha}.
\end{equation}
The area of the event
horizon is calculated as \begin{equation}  A =\int \sqrt{g_{\theta \theta} g_{\phi \phi}} d \theta d \phi =4 \pi r_\text{h}^2, \end{equation}
where the corresponding entropy of the black hole is $S=A/4$, and Hawking Temperature is
\begin{equation}
\begin{aligned}
 T_{\text{H}} &=\frac{F^{\prime}\left(r_\text{h}\right)}{4 \pi}=\frac{1}{8 \pi  m}-\frac{k}{7 m r_c \alpha}+\frac{2 \pi  \,k^{2}}{49 m \,r_c^{2} \alpha^{2}}.
\end{aligned}
\end{equation}

\subsection{Quantum tunneling of massive bosons from black hole} 
In this subsection, we study the quantum tunneling of massive bosons from the black hole to obtain the Hawking temperature of the black hole given in \eqref{metricc}. To do so, we use the Hamilton-Jacobi method to the tunneling approach \cite{Srinivasan:1998ty}, where one can see the event horizon as a potential barrier that we compute the probability of the tunneling particles from this potential by considering only the near the horizon and radial trajectories ( \(t - r\) plane).

To study the tunneling of massive bosons, we need to perturb the Klein-Gordon (KG) equation for the massive scalar particles defined as field \(\phi\) around a black hole geometry:
\begin{equation}
\hslash^2 g^{\mu\nu}\nabla_{\mu}\nabla_{\nu}\phi-m^{2}\phi=0,
\end{equation}
here \(m\) is the mass associated with the field \(\phi\).
Using the spherical harmonics decomposition, we write the KG equation for the black hole metric given in \eqref{metricc} as follows
\begin{equation}
\label{eqphi}
-\partial_{t}^{2}\phi+ F^{2}(r)\partial^{2}_{r}\phi+\frac{1}{2}\partial_{r}F^{2}(r)\partial_{r}\phi-\frac{m^{2}}{\hslash^{2}}F(r)\phi=0.
\end{equation} 
Then we use the Wentzel-Kramer-Brillouin (WKB) method by considering the field \(\phi\) as a semi-classical wave function of the particles to solve the equation \eqref{eqphi} with the help of ansatz 
\begin{equation}
\label{fi}
\phi(t,r)=\exp\left[-\frac{\imag}{\hslash}\mathcal{I}(t,r)\right].
\end{equation}
The \eqref{fi} is expanded for the lowest order in \(\hslash\):
\begin{equation} 
\label{hamilton-jacobi}
(\partial_{t}\mathcal{I})^{2}-F^{2}(r)(\partial_{r}\mathcal{I})^{2}-m^{2}F(r)=0.
\end{equation}

Note that Eq. \eqref{hamilton-jacobi} is the Hamilton-Jacobi equation with \(\mathcal{I}\) which has the role of relativistic action. We use the separation of variables: \eqref{hamilton-jacobi} 
\begin{equation}
\label{I}
\mathcal{I}(t,r) = -\omega t + W(r),
\end{equation}
where the \(\omega\) is the energy of the emitted particle. Then we solve the Eq. \eqref{hamilton-jacobi} by substituting Eq. \eqref{I} and find the spatial part of the action as follows: 
\begin{equation}
\label{eqW1}
W(r) = \int\frac{\dif r}{F(r)}\sqrt{\omega^2 - m^2F(r)}.
\end{equation}
To solve the above integral, we take an approximation of the function \(F(r)\) near the event horizon \(r_h\):
\begin{equation} 
F(r) = F(r_h) + F'(r_h)(r - r_h) + \cdots, 
\end{equation}
and the Eq. \eqref{eqW1} becomes

\begin{equation}
W(r) = \int \frac{\dif r}{F'(r_\text{h})}\frac{\sqrt{\omega^{2}-m^{2}F'(r_\text{h})(r-r_\text{h})}}{(r-r_\text{h})},
\end{equation}
where the prime stands for derivative relative to the radial coordinate.
To find the solution of the last integral, we use the residue theorem 
\begin{equation}
W = \frac{2\pi \imag\, \omega}{F'(r_\text{h})}.
\end{equation}

Hence the tunneling probability of a particle escape from the black hole is calculated as follows
\begin{equation}
\label{Gama}
\Gamma \sim \exp(-2\,\Im(\mathcal{I})) = \exp\bigg[-\frac{4\pi\omega}{F'(r_\text{h})}\bigg],
\end{equation}
where we note that \(\Im\mathcal{I} = \Im W\).
If you compare Eq. \eqref{Gama} with the Boltzmann factor \(e^{-\omega/T}\), it is easy to see that the Hawking temperature of the black hole is
\begin{equation}
\label{Th}
T_{\text{H}} = \frac{\omega}{2\,\Im(\mathcal{I})} = \frac{F'(r_\text{h})}{4\pi}.
\end{equation}
Using the event horizon in \eqref{hor},
the Hawking temperature is founded as
\begin{equation}
\label{eq:temperature}
 T_{\text{H}} = \frac{F^{\prime}\left(r_\text{h}\right)}{4 \pi}=\frac{1}{8 \pi  m}-\frac{k}{7 m r_c \alpha}+\frac{2 \pi  \,k^{2}}{49 m \,r_c^{2} \alpha^{2}}.
\end{equation}
Note that in the limit $k\to 0$ we recover the usual Schwarzschild temperature for the classical black hole ($T_\text{Sch}=1/4\pi r_\text{h}$, assuming that $r_\text{h}=2m$ is the Schwarzschild radius).

\section{Shadow cast and Thin-accretion disk of Quantum Wave Dark Matter Black hole} \label{sec3}
In this section, we will now study the behavior of the shadow radius due to the free parameters $m_\text{b}$ and $r_\text{c}$. Our first goal is to constraint the value of the soliton core radius $r_\text{c}$ that envelopes the black hole Sgr. A* and M87* using a certain range of boson mass: $10^{-23}\text{ eV}\leq m_\text{b}\leq 10^{-21}\text{ eV}$. According to Schive \citep{Schive:2014dra}, if $m_\text{b} = 10^{-22}$ eV, then the expected soliton mass and core radius are $1.44\times10^9 M_\odot$ and $160$ pc, respectively. Note that these values are interpreted as the \textit{maximum} value for the soliton mass, and the \textit{minimum} value for the core radius \cite{Li:2020qva} since the soliton core might be further developing \cite{Hui:2016ltb}. The model also did not consider any black hole at the galactic center. Here, we will consider such a scenario.

\begin{figure}
    \centering
    \includegraphics[width=0.48\textwidth]{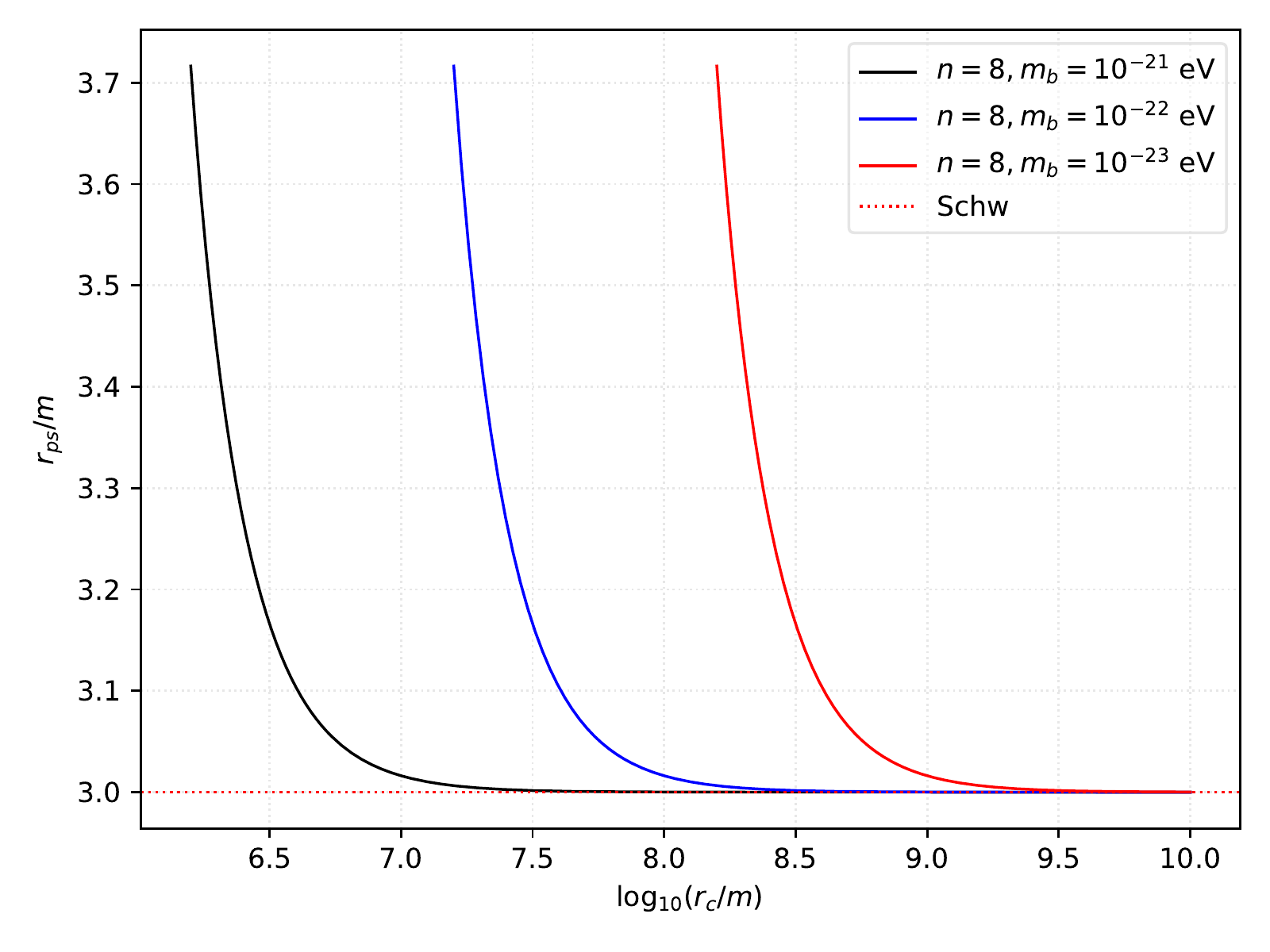}
    \caption{Behavior of photonsphere under the effect of the fuzzy dark matter.}
    \label{rph}
\end{figure}

To begin, consider the Lagrangian for null geodesic:
\begin{equation} \label{e22}
	\mathcal{L} = \frac{1}{2}\left( -A(r) \dot{t} +B(r) \dot{r} + C(r) \dot{\phi} \right).
\end{equation}
After using the Euler-Lagrange variational principle, we obtain two constants of motion
Constants of motion
\begin{equation} \label{e23}
    E = A(r)\frac{dt}{d\lambda}, \qquad L = C(r)\frac{d\phi}{d\lambda},
\end{equation}
where the ratio $L/E$ is defined as the impact parameter.
\begin{equation} \label{e24}
    b \equiv \frac{L}{E} = \frac{C(r)}{A(r)}\frac{d\phi}{dt}.
\end{equation}
$ds^2 = 0$ is required for null particles, which allows us to obtain the orbit equation
\begin{align} \label{e25}
    \left(\frac{dr}{d\phi}\right)^2 =\frac{C(r)}{B(r)}\left(\frac{h(r)^2}{b^2}-1\right),
\end{align}
where $h(r)$ is defined as \cite{Perlick:2015vta}
\begin{equation} \label{e26}
    h(r)^2 = \frac{C(r)}{A(r)}.
\end{equation}
Such a function is so useful since the condition $h'(r) = 0$ will allow us to obtain an expression that will extract the value of the photonsphere radius $r_\text{ps}$. Our result is
\begin{align} \label{e27}
	&\Bigg\{k \left[\frac{\alpha^{n} \lambda \left(2 n -1\right) r^{2+\sigma} \pi^{\frac{5}{2}}}{8}+\left(\frac{r_\text{c}^{2+\sigma} h_2 \sigma \sqrt{\alpha}}{2}-\frac{h_1 \left(\sigma +1\right) \left(2 n -1\right) r_\text{c}^{\sigma} r^{2} \alpha^{\frac{3}{2}}}{4}\right) \pi  N \right] -\frac{\lambda r^{3+\sigma} \left(2 n -1\right) N \alpha^{\frac{3}{2}+n}}{8}\Bigg\} \times \nonumber \\
	&\exp \left[\frac{4 r^{-\sigma -1} k \,\alpha^{-n -\frac{3}{2}}}{\lambda N} r_\text{c}^{\sigma} \pi  \,\alpha^{\frac{3}{2}} h_1 N -\frac{\alpha^{n} r^{\sigma} \pi^{\frac{5}{2}} \lambda}{2} \right] + \frac{3 m \,\alpha^{\frac{3}{2}+n} \lambda r^{2+\sigma} \left(2 n -1\right) N}{8} = 0
\end{align}
where, in addition, we wrote another hypergeometric function as
\begin{equation} \label{e28}
    h_2 = {}_2F_{1}\left(n,n-\frac{1}{2};n+\frac{1}{2};-\frac{r_\text{c}^2}{r^{2} \alpha} \right).
\end{equation}
From the looks of it, Eq. \eqref{e27} is a bit of a worked-out equation. We cannot extract simply an analytical formula for $r_\text{ps}$. Hence, we rely on numerical analysis, where the results are plotted in Fig. \ref{rph}. The general trend of the curves are the same for different boson mass and we can say that there is a certain minimum for $r_\text{c}$ for each $m_\text{b}$. Such a minimum is larger for lower boson mass. As the core radius increases, $r_\text{ps} \to 3m$ which is the Schwarzschild case.

\begin{figure*}
    \centering
    \includegraphics[width=0.48\textwidth]{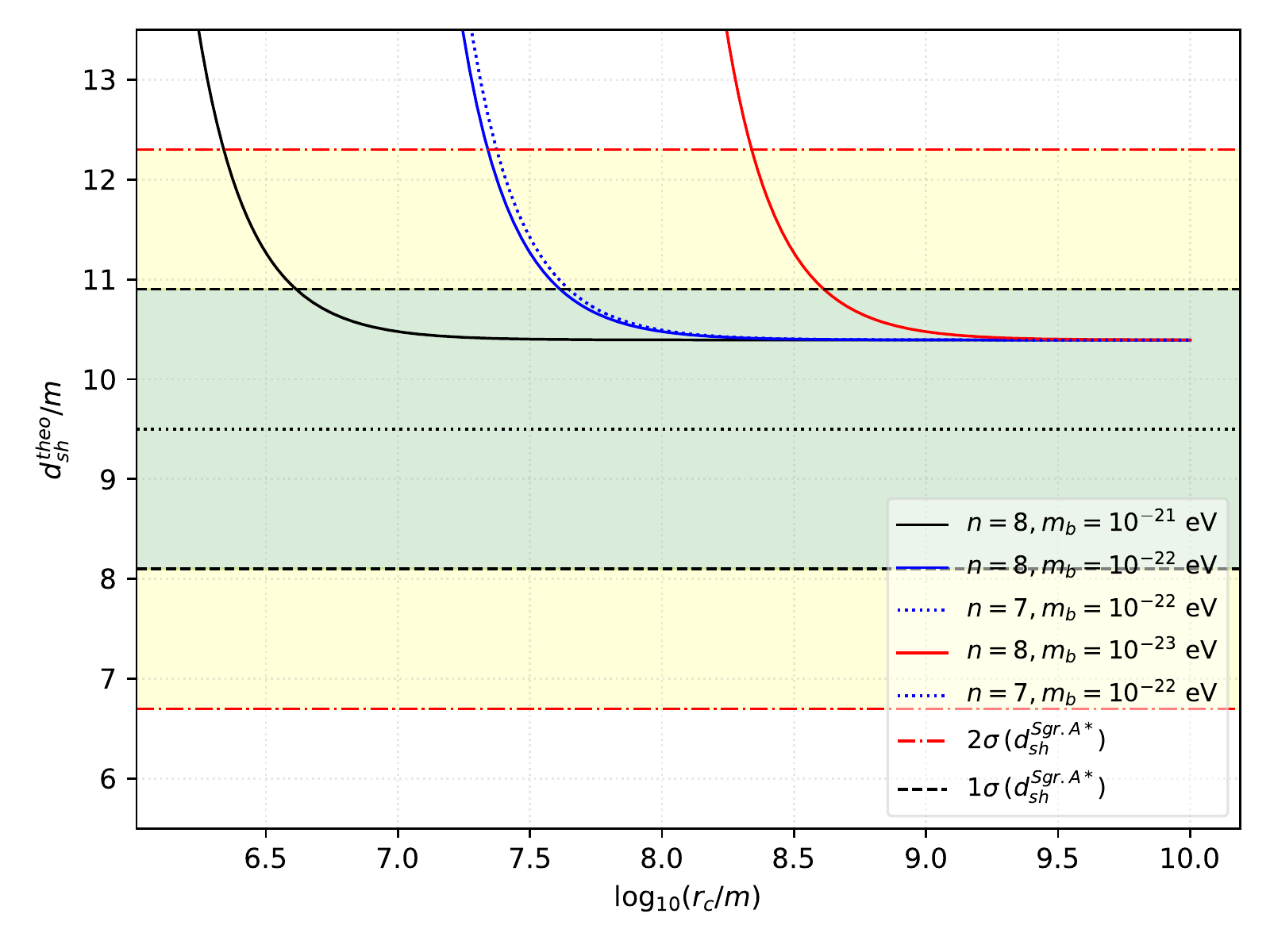}
    \includegraphics[width=0.48\textwidth]{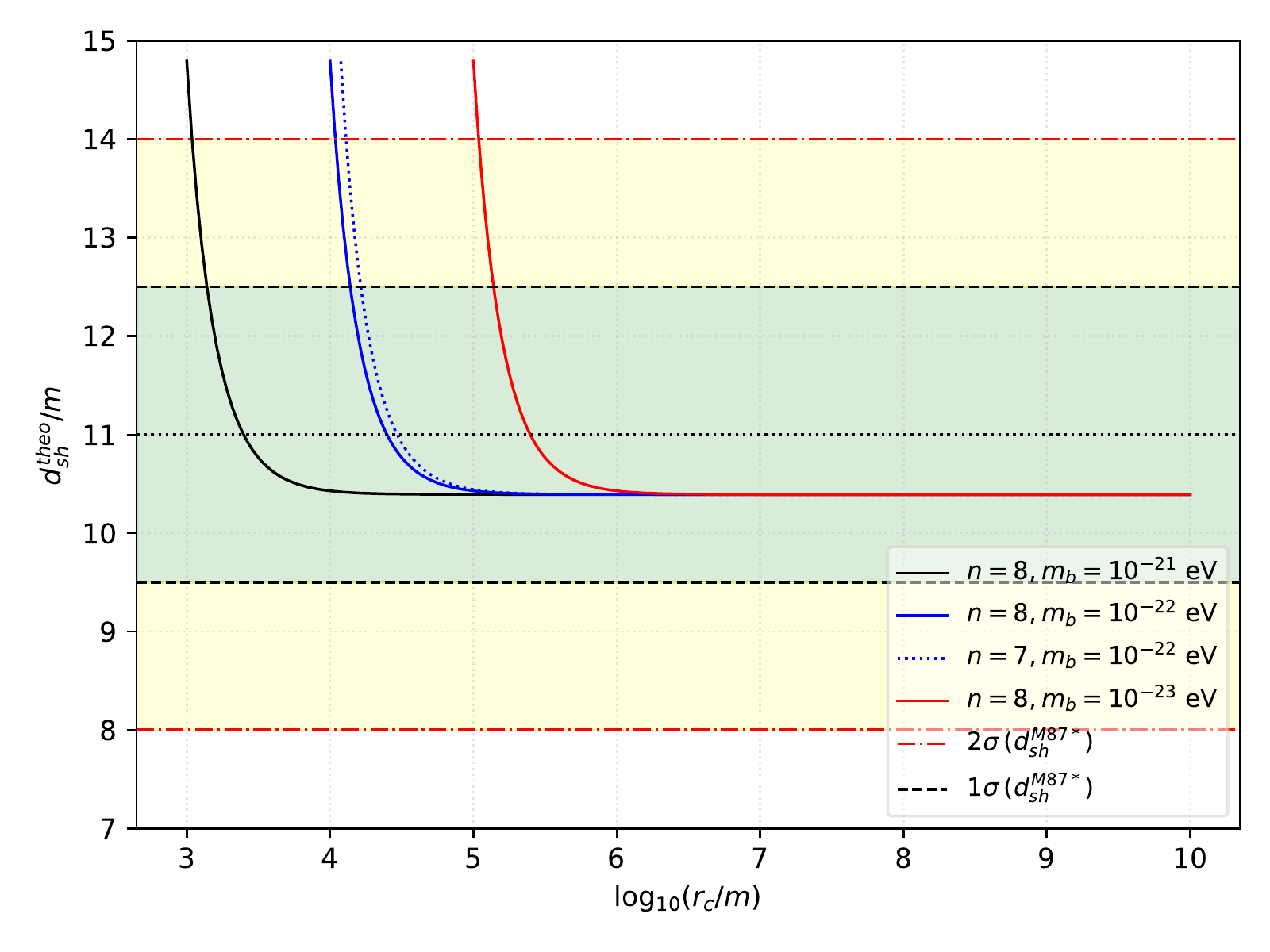}
    \caption{Left: Sgr. A*. Right: M87*.}
    \label{shacons}
\end{figure*}

Next, consider an observer located at some finite distance $r_\text{o}$ from the black hole. Simple geometrical construction will allow us to define
\begin{equation} \label{e29}
    \tan(\alpha_{\text{sh}}) = \lim_{\Delta x \to 0}\frac{\Delta y}{\Delta x} = \left(\frac{C(r)}{B(r)}\right)^{1/2} \frac{d\phi}{dr} \bigg|_{r=r_\text{o}},
\end{equation} 
and can be recast as
\begin{equation} \label{e30}
    \sin^{2}(\alpha_\text{sh}) = \frac{b_\text{crit}^{2}}{h(r_\text{o})^{2}}.
\end{equation}
Here, the critical impact parameter which is a function of $r_\text{ps}$ helps define the shadow contour and should be obtained first. Note that in the pure Schwarzschild case, the shadow radius $R_\text{sh}$ is equal $b_\text{crit}$. However, this is not always the case since, for example, in a non-asymptotically flat spacetime such as the Kottler spacetime, $R_\text{sh} \ne b_\text{crit}$. The condition $\frac{d^2r}{d\phi^2} = 0$ gives a useful formula to derive the critical impact parameter \cite{Pantig:2022ely}:
\begin{equation}
    b_\text{crit}^2 = \frac{h(r_\text{ps})}{\left[B'(r_\text{ps})C(r_\text{ps})-B(r_\text{ps})C'(r_\text{ps})\right]} \Bigg[h(r_\text{ps})B'(r_\text{ps})C(r_\text{ps})-h(r_\text{ps})B(r_\text{ps})C'(r_\text{ps})-2h'(r_\text{ps})B(r_\text{ps})C(r_\text{ps}) \Bigg].
\end{equation}
which gives
\begin{align} \label{e31}
	&b_\text{crit}^2 = 4 N r_\text{ps}^{5} \lambda \alpha^{\frac{5}{2}} \left(n -\frac{1}{2}\right) \Bigg\{ \Bigg\{4 \pi  r_\text{c}^{2+\sigma} k \lambda r_\text{ps}^{-\sigma} \alpha^{-\frac{\sigma}{2}} N h_2 - 2 r_\text{c}^{\sigma} \pi  k \alpha^{-n +\frac{5}{2}} \left(2 n -1\right) r_\text{ps}^{-\sigma +2} \left(\sigma +1\right) N h_1 \nonumber \\
	&+\left(k \pi^{\frac{5}{2}} \alpha +r_\text{ps} N \alpha^{\frac{5}{2}}\right) r_\text{ps}^{2} \lambda \left(2 n -1\right) \Bigg\} \exp\Bigg\{\frac{4 \pi  k}{\alpha^{\frac{3}{2}} \lambda N r_\text{ps}} \left[ \alpha^{-\frac{\sigma}{2}} \left(\frac{r_\text{c}}{r_\text{ps}}\right)^{\sigma} h_1 N -\frac{\pi^{\frac{3}{2}} \lambda}{2}
\right] \Bigg\}-m r_\text{ps}^{2} \alpha^{\frac{5}{2}} \lambda \left(2 n -1\right) N \Bigg\} ^{-1}.
\end{align}
Finally, since $D(r) = C(r)$, and using Eq. \eqref{e30}, we simply obtain the shadow radius as
\begin{equation} \label{e32}
	R_\text{sh} = b_\text{crit}\sqrt{A(r_\text{o})},
\end{equation}
where we should use Eqs. \eqref{e27} and \eqref{e31}.
\begin{figure*}
    \centering
    \includegraphics[width=0.30\textwidth]{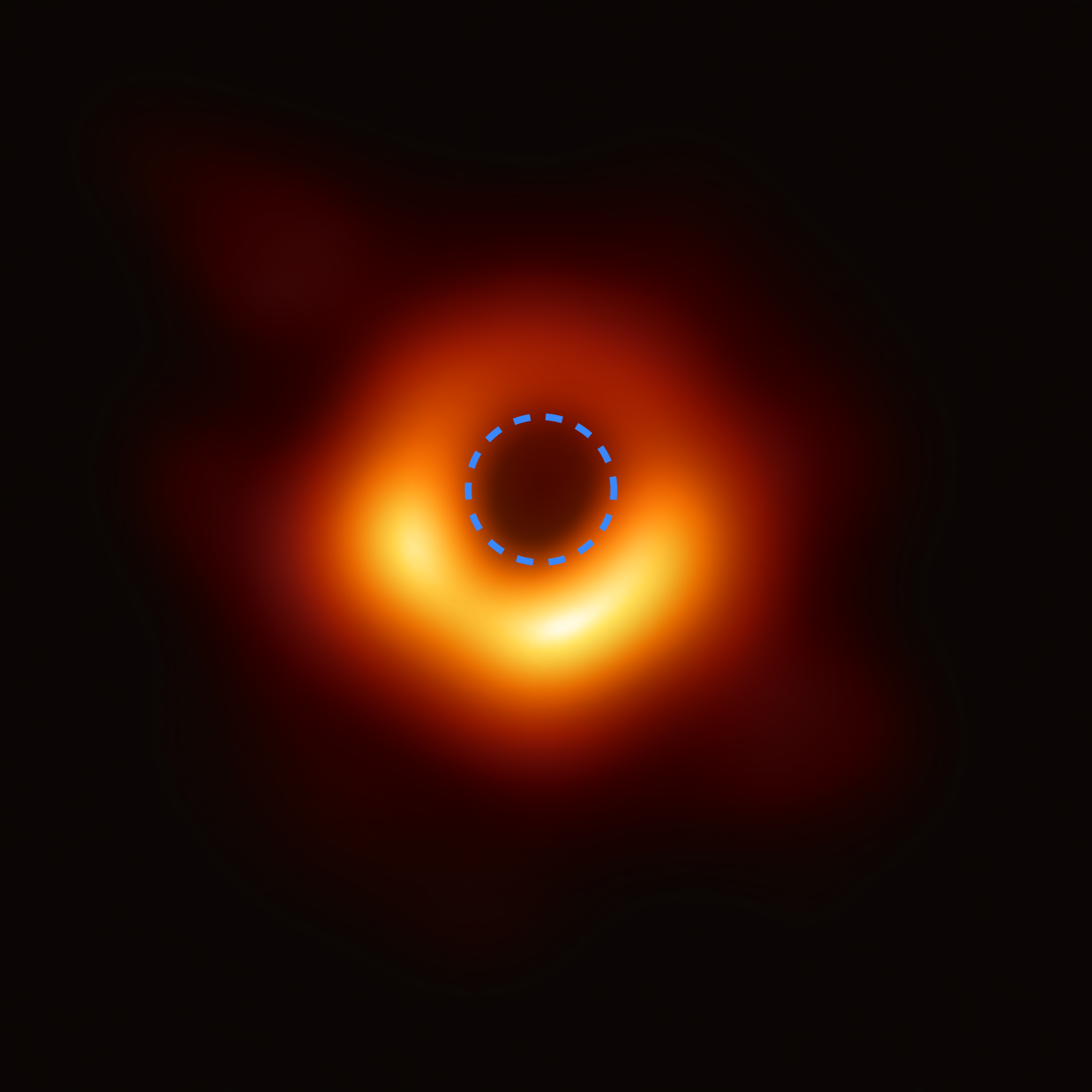}
    \includegraphics[width=0.30\textwidth]{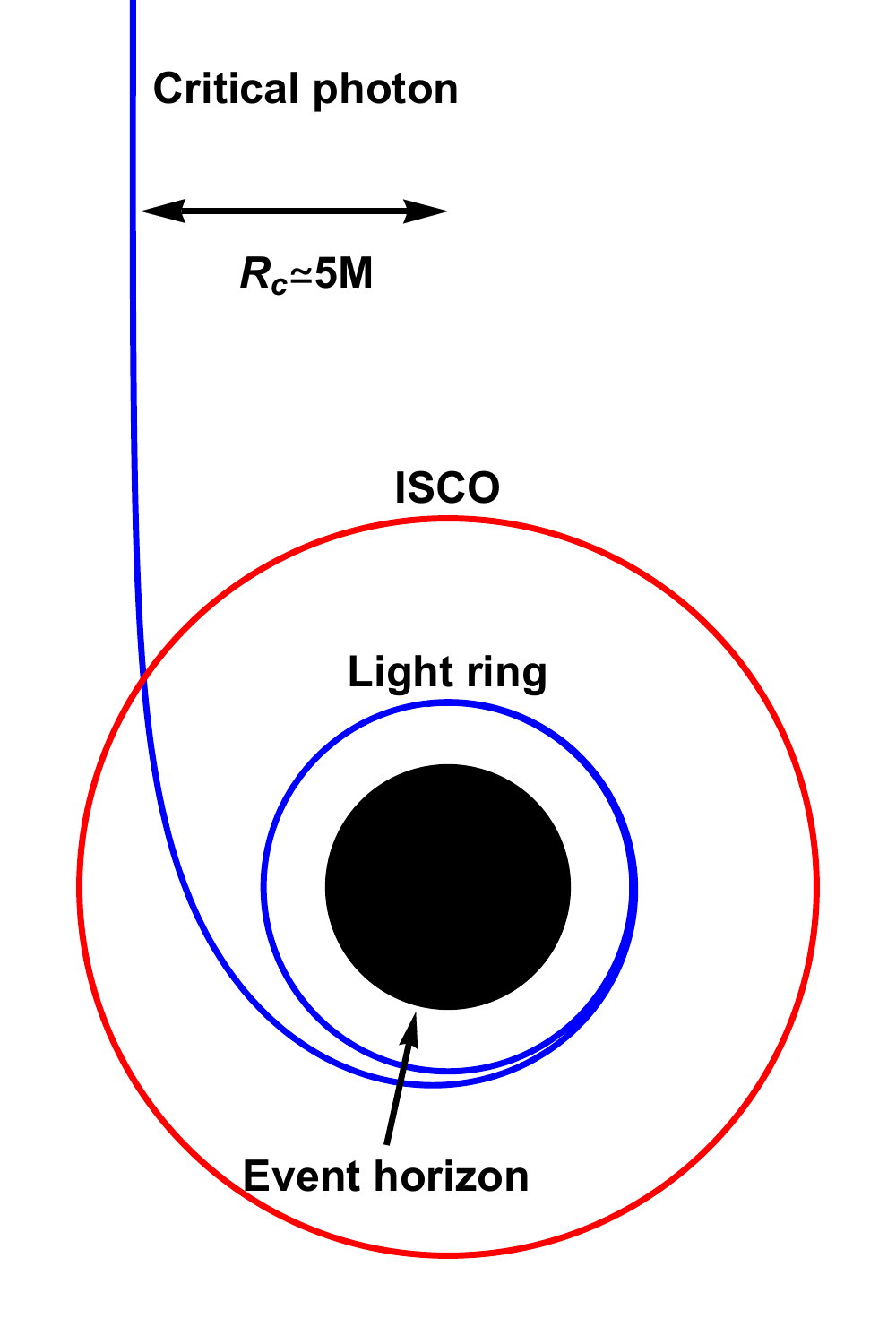}
    \caption{Credit: Event Horizon Telescope Collaboration}
    \label{fig:my_label}
\end{figure*}
We use Eq. \eqref{e32} to constrain $r_\text{c}$ using the EHT data. We summarize the observed data in Table \ref{tab1} (see Refs. \cite{EventHorizonTelescope:2019dse,EventHorizonTelescope:2022xnr}).
\begin{table}
    \centering
    \begin{tabular}{ p{2cm} p{3.5cm} p{4.5cm} p{2cm}}
    \hline
    \hline
    Black hole & Mass $m$ ($M_\odot$) & Angular diameter: $2\alpha_\text{sh}$ ($\mu$as) & Distance (kpc) \\
    \hline
    Sgr. A*   & $4.3 \pm 0.013$x$10^6$ (VLTI)    & $48.7 \pm 7$ (EHT) &   $8.277 \pm 0.033$ \\
    M87* &   $6.5 \pm 0.90$x$10^9$  & $42 \pm 3$   & $16800$ \\
    \hline
    \end{tabular}
    \caption{Black hole observational constraints.}
    \label{tab1}
\end{table}
We can then use the formula
\begin{equation} \label{e33}
    d_\text{sh} = \frac{D \theta}{m},
\end{equation}
which gives the following values for the diameter of the shadow of M87* and Sgr. A*. These are $d^\text{M87*}_\text{sh} = (11 \pm 1.5)m$, and $d^\text{Sgr. A*}_\text{sh} = (9.5 \pm 1.4)m$, respectively. On the other hand, our black hole model's theoretical shadow diameter can be easily found with $d_\text{sh}^\text{theo} = 2R_\text{sh}$. We plot the result in Fig. \ref{shacons}, and Table \ref{tab2} summarizes the upper bounds in both $1-$ and $2\sigma$ levels.
\begin{table}
    \centering
    \begin{tabular}{c| c c c}
    \hline
    \hline
    Sgr. A* & \multicolumn{3}{c}{Upper bound $\log_{10}(r_\text{c}/m)$} \\
    \hline
    $\sigma$ level & $10^{-21}$ & $10^{-22}$ & $10^{-23}$   \\
    \hline
    $1\sigma$  & $6.61$ & $7.61$ & $8.61$  \\
    $2\sigma$  & $6.34$ & $7.34$ & $8.34$ \\
    \hline
    \end{tabular}
    \quad
    \begin{tabular}{c|c c c}
    \hline
    \hline
    M87* & \multicolumn{3}{c}{Upper bound $\log_{10}(r_\text{c}/m)$} \\
    \hline
    $\sigma$ level & $10^{-21}$ & $10^{-22}$ & $10^{-23}$   \\
    \hline
    $1\sigma$  & $3.14$ & $4.14$ & $5.14$  \\
    $2\sigma$  & $3.04$ & $4.04$ & $5.04$ \\
    \hline
    \end{tabular}
    \caption{The upper bound values of the soliton core $r_\text{c}/m$ as depicted in Fig. \ref{shacons}.}
    \label{tab2}
\end{table}
As mentioned earlier, without taking into consideration the black hole at the galactic center, the expected minimum soliton core radius is ranged from $160-180$ pc when $m_\psi = 1$ for DM halo mass ranging from $M_\text{h} \sim 1.5-2.0\times10^{12} M_\odot$ \cite{Schive:2014dra,Li:2020qva}. If we take the average, $r_\text{c} = 170$ pc corresponds to $\log_{10}(r_\text{c}/m) \sim 8.92$ for Sgr. A*. If we can observe the same behavior of the shadow deviation for M87*, then it can be estimated that $\log_{10}(r_\text{c}/m) \sim 5.70$ which corresponds to a soliton core radius of $r_\text{c} = 156$ pc. We should note that at such a minimum value, the deviation in the shadow radius caused by the soliton profile is nearly the same as the Schwarzschild case. However, as the soliton core radius lessens up to the limit imposed by the confidence levels, we can observe some drastic increase in the shadow radius. Thus, when the black hole is considered, the minimum value for $r_\text{c}$ is further lessened and constrained. We could also see the effect of the boson mass. That is, we observe that when $m_\text{b}$ increases, the minimum value for the required $r_\text{c}$  decreases. Hence, the dark matter made of soliton becomes more concentrated. Finally, we also considered $n=7$ in the plot for $m_\psi = 1$ only since the same conclusion can be made for other values of $m_\psi$. Results indicate that as the value of $n$ decreases, the trend is to increase the constrained value for $r_\text{c}$. However, as emphasized earlier, setting a different value for $n$ may provide results that are ruled out by observation using other astrophysical simulations.

The result for M87* is also worth discussing. While the general trend is the same as in Sgr. A*, we noticed that there is a certain value for $r_\text{c}$ that produces a shadow radius that is similar to the observed mean, which is $d_\text{sh}=11m$. For $m_\psi = 1$, this is $r_\text{c} \sim 7.81$ pc ($12560r_\text{h}$). Indeed, this is much smaller soliton core radius, and somehow consistent with the estimate using a different methodology in Ref. \cite{Davies:2019wgi} where $r_\text{c} = 10$ pc.

\begin{figure*}
    \centering
    \includegraphics[width=0.48\textwidth]{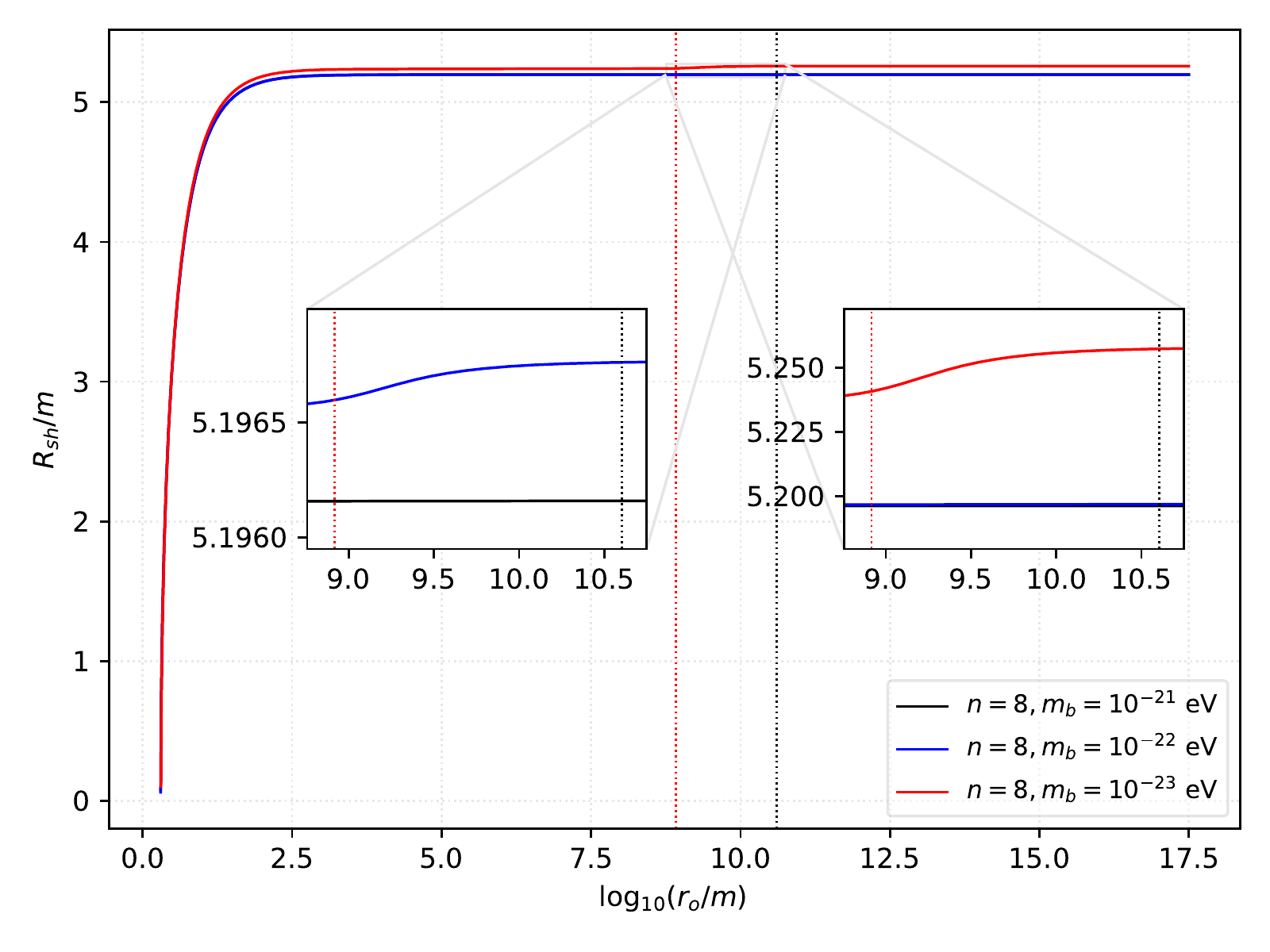}
    \includegraphics[width=0.48\textwidth]{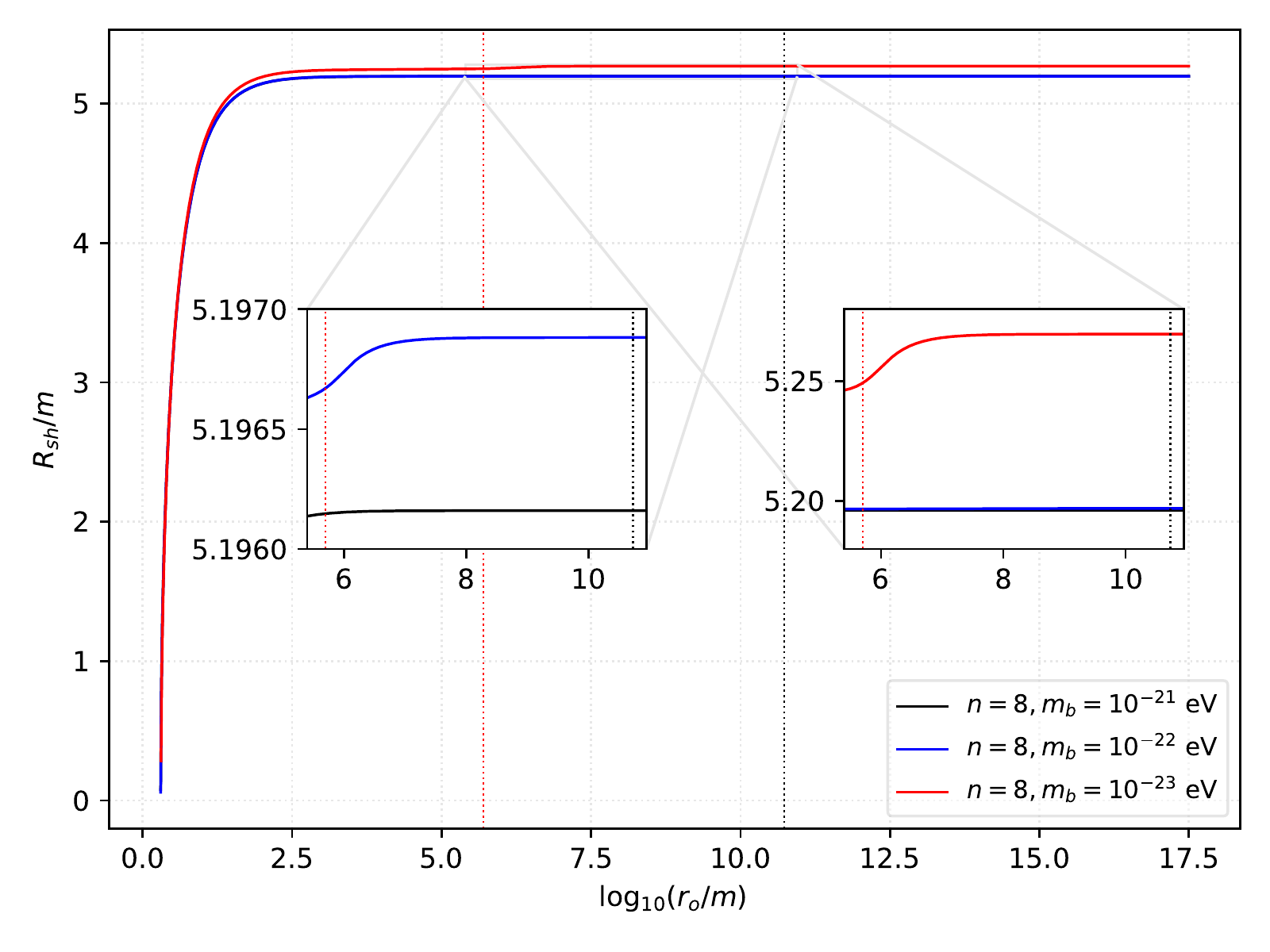}
    \caption{The behavior of the shadow radius due to a static observer at different locations $r_\text{o}$. Left: Sgr. A*. Right: M87. The red and black vertical dotted lines correspond to $r_\text{c}$ and the actual distance of Earth from the SMBHs, respectively.}
    \label{shadow}
\end{figure*}

Now that we know the effect of the soliton profile on the shadow radius, let us pick some certain values of $r_\text{c}$ within the confidence intervals. For simplicity, let us take $r_\text{c} = 170$ pc for Sgr. A*, and $r_\text{c} = 156$ pc for M87*. Looking at Fig, \ref{shacons}, we expect that the deviation would be small for these soliton cores, at least when the static observer is at $r \to \infty$. In Fig. \ref{shadow}, we plot what would happen to the shadow radius if the observer is inside or outside the soliton core. We can see that through the inset plots, the shadow radius increases as the observer passes outside the soliton core radius. Interestingly, the lower the boson mass, the greater the deviation is seen. Nevertheless, the general trend of how the shadow radius behaves is the same for both Sgr. A* and M87*

\subsection{Spherically infalling accretion}
Here we investigate the realistic visualization of the shadow cast with the spherically free falling accretion disk model around the black hole, similar way with \cite{Jaroszynski:1997bw,Bambi:2012tg}. For this purpose, we use the specific-intensity observed at the observed photon frequency $\nu_\text{obs}$ by obtaining the integral along the light ray:
        \begin{equation}
            I(\nu_\text{obs},b_\gamma) = \int_\gamma g^3 j(\nu_e) dl_\text{prop},
            \label{eq:bambiI}
        \end{equation}
        
where the impact parameter is $b_{\gamma}$, and the emissivity/volume is $j(\nu_e)$. Moreover,  $dl_\text{prop}$ is for the infinitesimal proper length and $\nu_e$ is for the photon frequency of the emitter. Define the redshift-factor for the infalling accretion:
        \begin{equation}
            g = \frac{k_\mu u^\mu_o}{k_\mu u^\mu_e},
        \end{equation}
in which the four-velocity of the photon is $k^\mu=\dot{x}_\mu$ and four-velocity of the distant observer is $u^\mu_o=(1,0,0,0)$. Next,  we write the four-velocity of the infalling accretion $u^\mu_e$
\begin{equation}
u_{\mathrm{e}}^{t}=\frac{1}{A(r)}, \quad u_{\mathrm{e}}^{r}=-\sqrt{\frac{1-A(r)}{A(r) B(r)}}, \quad u_{\mathrm{e}}^{\theta}=u_{\mathrm{e}}^{\phi}=0.
\end{equation}
Then we write the constant of the photon motion with relation  $k_{\alpha} k^{\alpha}=0$, to obtain $k_{r}$ and $k_{t}$:
\begin{equation}
k_{r}=\pm k_{t} \sqrt{B(r)\left(\frac{1}{A(r)}-\frac{b^{2}}{r^{2}}\right)}.
\end{equation}
Here, the sign $\pm$ stands for the photon
approaching or moving away to/from the black hole. Afterward, we write redshift factor $g$ and proper distance $dl_\gamma$
   \begin{equation}
   g = \Big( u_e^t + \frac{k_r}{k_t}u_e^r \Big)^{-1},
  \end{equation}
  and
 \begin{equation}
  dl_\gamma = k_\mu u^\mu_e d\lambda = \frac{k^t}{g |k_r|}dr.
\end{equation}

For specific emissivity, we use the monochromatic emission with a frequency of rest frame $\nu_*$ as follows:
        \begin{equation}
            j(\nu_e) \propto \frac{\delta(\nu_e - \nu_*)}{r^2},
        \end{equation}

the equation of intensity in \eqref{eq:bambiI} reduces to
        
        \begin{equation}
            F(b_\gamma) \propto \int_\gamma \frac{g^3}{r^2} \frac{k_e^t}{k_e^r} dr.
        \end{equation}

\begin{figure*}
    \centering
    \includegraphics[width=0.48\textwidth]{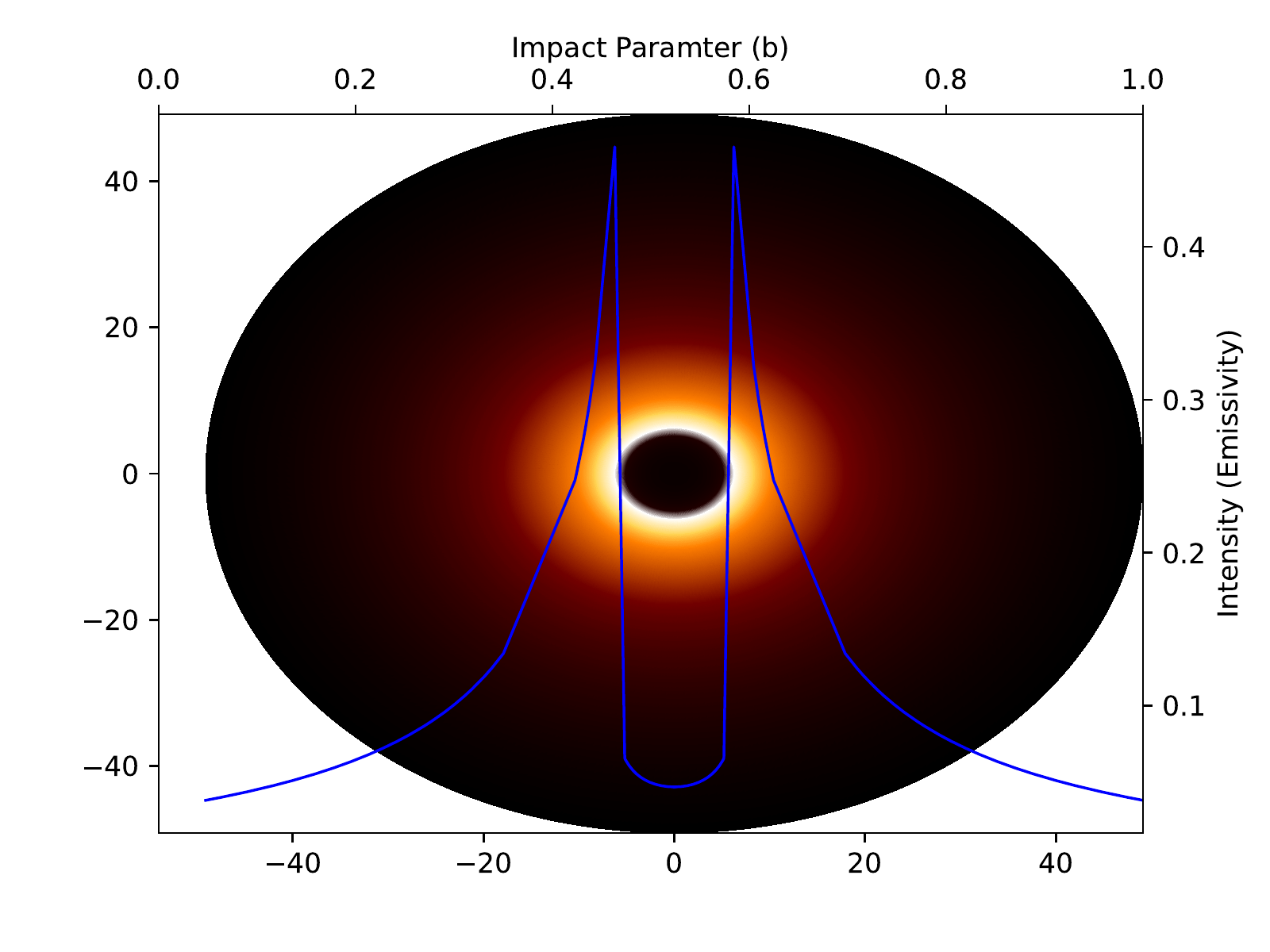}
    \includegraphics[width=0.48\textwidth]{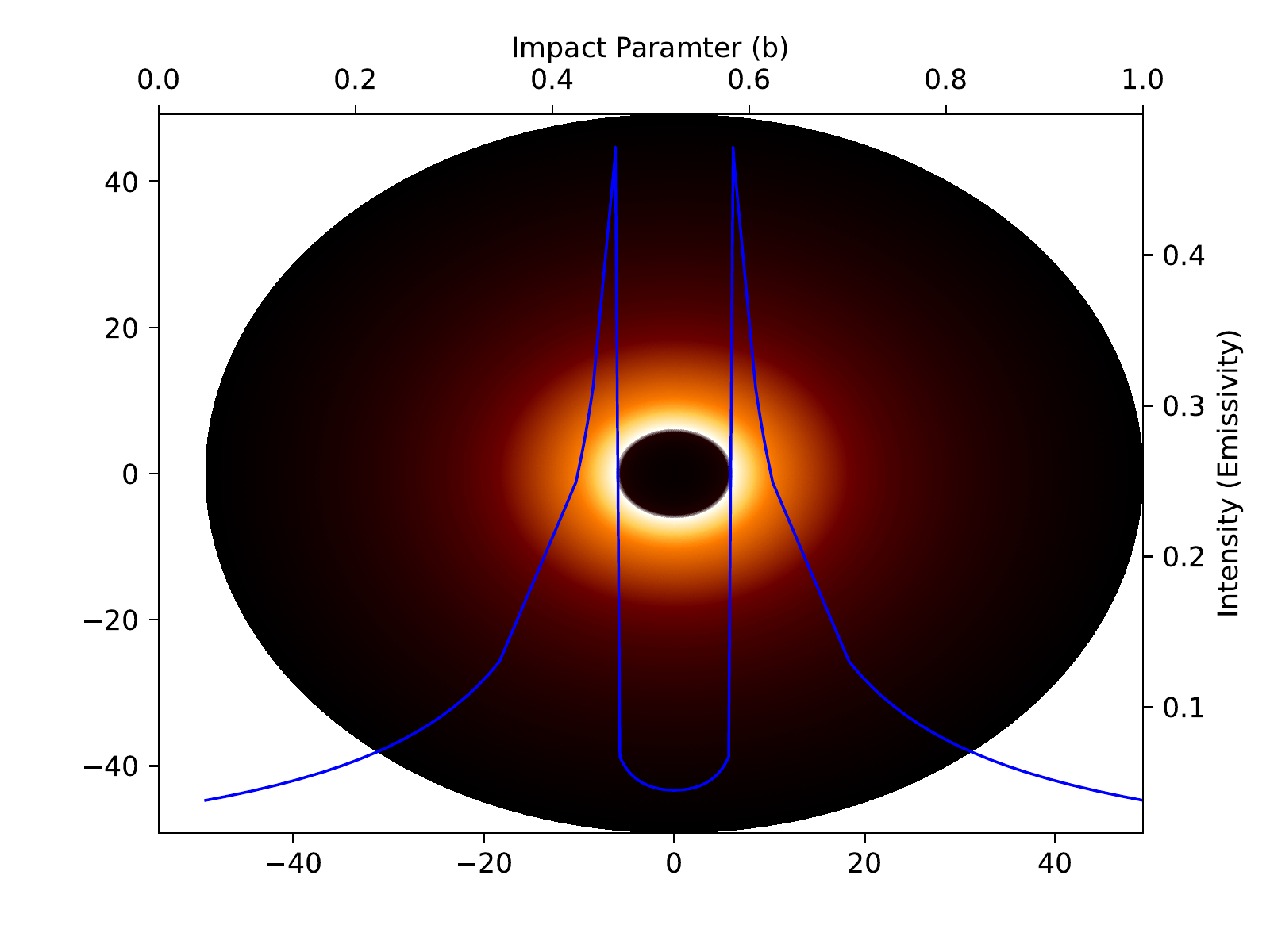}
    \caption{Left: Schwarzschild case. Right: Shadow cast and specific intensity $I_{obs}$ as seen by a distant observer for M87*. Here, $m_\text{b} = 1$ and we used $r_\text{c} = 7.81$ pc (corresponding to $d^\text{theo}_\text{sh} = 11m$ in Fig. \ref{shacons}).}
    \label{fig:thinacc1}
\end{figure*}

\begin{figure*}
    \centering
    \includegraphics[width=0.48\textwidth]{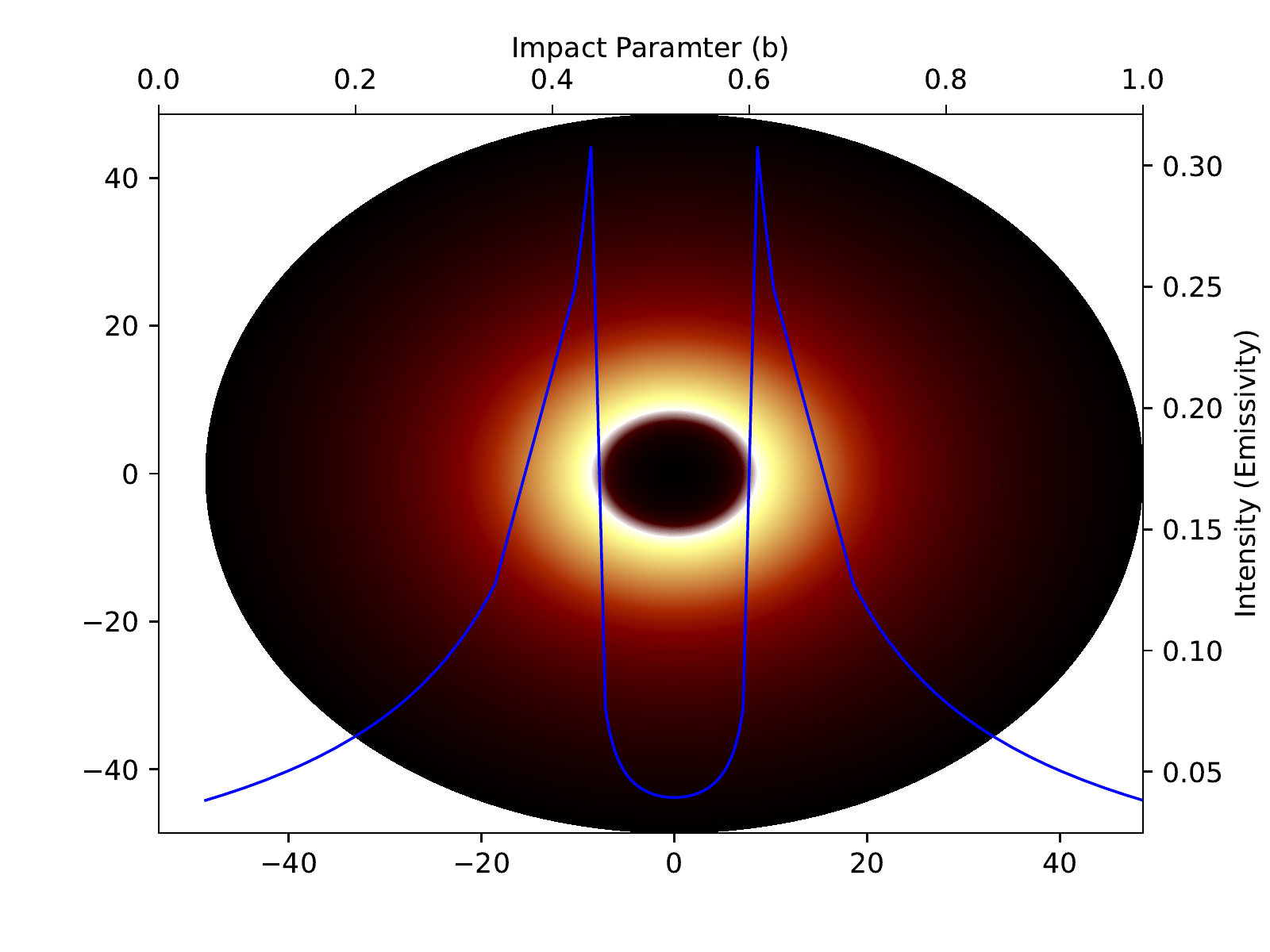}
    \includegraphics[width=0.48\textwidth]{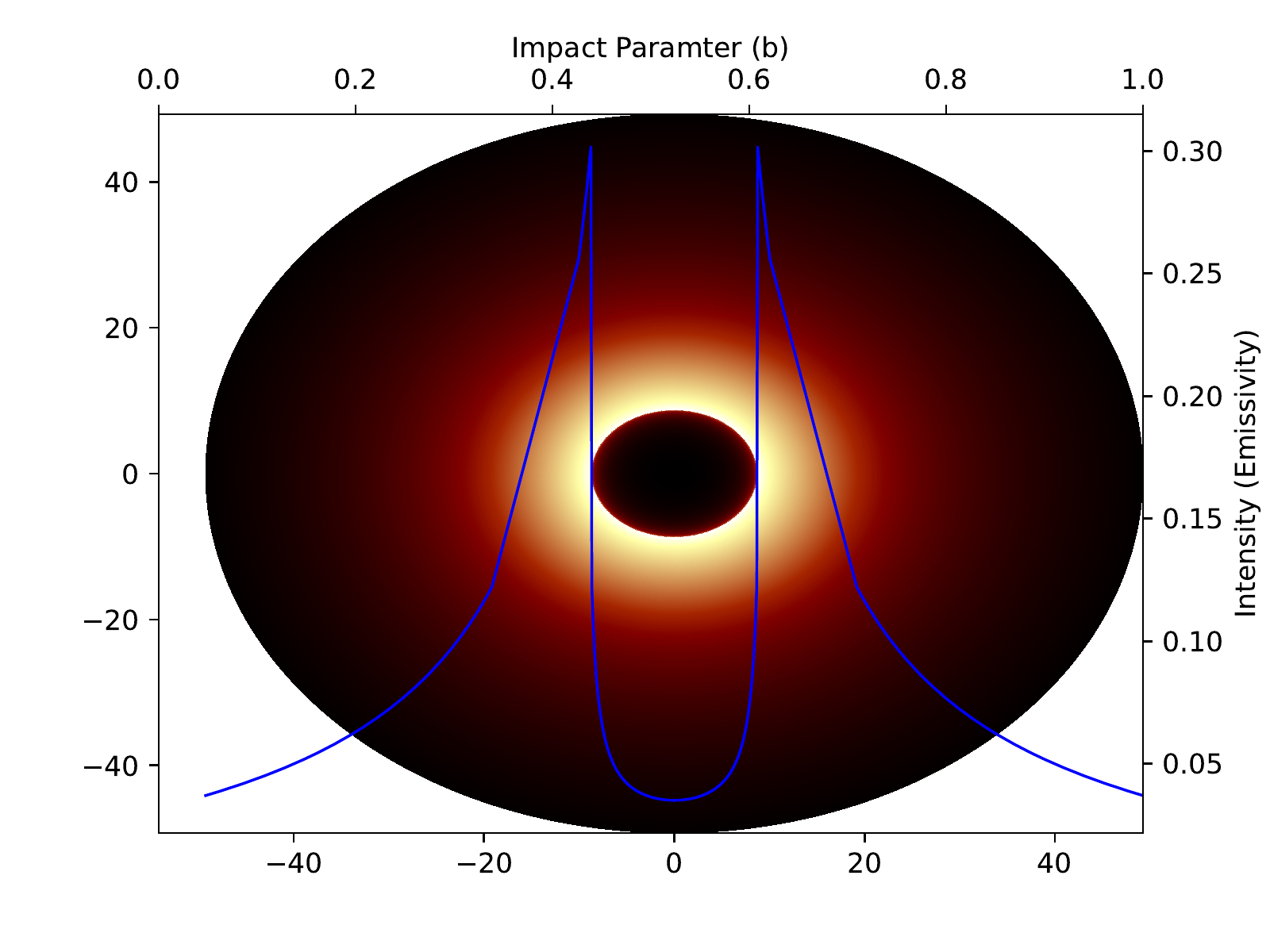}
    \caption{Left: $r_\text{c} = 4.29$ pc for M87*. Right: $r_\text{c} = 3.41$ pc for M87*. In both panels, we used $m_\text{b} = 1$. These values for $r_\text{c}$ corresponds to the result shown in Table \ref{tab2}.}
    \label{fig:thinacc2}
\end{figure*}

To study the shadow with the accretion disk, one should solve the above equation. We solve it numerically using EinsteinPy similarly with \cite{Bapat:2020xfa,Okyay:2021nnh,Chakhchi:2022fls,Kuang:2022xjp,Uniyal:2022vdu,Pantig:2022ely}. It gives us the flux which shows the effects of the quantum wave dark matter on the specific intensity seen by a distant observer for an infalling accretion in Figs. (\ref{fig:thinacc1}, and \ref{fig:thinacc2}). 

In Figs. \ref{fig:thinacc1}-\ref{fig:thinacc2}, we overlap the intensity plot to the black hole shadow cast. Here, the brightest is the photon ring (represented by the peak of the intensity curve). Using the constraints in Table \ref{tab2}, we see that there is no discerning difference occurs between the Schwarzschild case and the soliton case where $m_\text{b} = 1$. However, when we use the value of $r_\text{c}$ for $1\sigma$ and $2\sigma$ confidence levels, we see a slight increase in the shadow size while the peak intensity decreases. In Fig. \ref{fig:thinacc2}, we can also notice a faint luminosity of light near the contour of the event horizon, which can be attributed to the soliton dark matter effect as the photon travels through such an astrophysical environment.

\subsection{Weak photon deflection using the Gauss-bonnet theorem}
In this section, we will probe the parameter $\epsilon$ and $\Lambda_0$ using the weak deflection angle $\hat{\alpha}$. With this aim, we use the Gauss-Bonnet theorem, which states that \cite{Carmo2016,Klingenberg2013}
\begin{equation} \label{eGBT}
    \iint_M KdS+\sum\limits_{i=1}^N \int_{\partial M_{a}} \kappa_{\text{g}} d\ell+ \sum\limits_{i=1}^N \theta_{i} = 2\pi\chi(M),
\end{equation}
where $K$ is the Gaussian curvature, $dS$ is the area measure, $\theta_i$, and $\kappa_g$ is the jump angles and geodesic curvature of $\partial M$, respectively, and $d\ell$ is the arc length measure. Its application to null geodesic at the equatorial plane implies that the Euler characteristics should be $\chi(M) = 1$. If the integral is evaluated over the infinite area surface bounded by the light ray, it was shown \cite{Ishihara:2016vdc} that the above reduces to
\begin{equation} \label{eIshi}
	\hat{\alpha}=\phi_{\text{RS}}+\Psi_{\text{R}}-\Psi_{\text{S}} = -\iint_{_{\text{R}}^{\infty }\square _{\text{S}}^{\infty}}KdS,
\end{equation}
where $\hat{\alpha}$ is the weak deflection angle. In the above formula, $\phi_{RS} = \Psi_R - \Psi_S$ is the azimuthal separation angle between the source S and receiver R, $\Psi_R$ and $\Psi_S$ are the positional angles, and $_{R}^{\infty }\square _{S}^{\infty}$ is the integration domain. It was shown in \cite{Li:2020wvn} that if one uses the path in the photonsphere orbit instead of the path at infinity, the above can be recast in a form applicable for non-asymptotically flat spacetimes:
\begin{equation} \label{eLi}
    \hat{\alpha} = \iint_{_{r_\text{co}}^{R }\square _{r_\text{co}}^{S}}KdS + \phi_{\text{RS}}.
\end{equation}

To determine $K$ and $dS$, consider that metric from an SSS spacetime
\begin{align}
    ds^2=g_{\mu \nu}dx^{\mu}dx^{\nu}=-A(r)dt^2+B(r)dr^2+C(r) d\theta^2 + D(r) \sin^2\theta d \phi^2.
\end{align}
Due to spherical symmetry of the metric, it will suffice to analyze the deflection angle when $\theta = \pi/2$, thus, $D(r) = C(r)$. Since we are also interested in the deflection angle of massive particles, we need the Jacobi metric, which states that
\begin{align} \label{eJac}
    dl^2=g_{ij}dx^{i}dx^{j}
    =(E^2-\mu^2A(r))\left(\frac{B(r)}{A(r)}dr^2+\frac{C(r)}{A(r)}d\phi^2\right),
\end{align}
where the energy per unit mass of the massive particle is
\begin{equation} \label{en}
    E = \frac{\mu}{\sqrt{1-v^2}}.
\end{equation}
It is then useful to define another constant quantity in terms of the impact parameter $b$, which is the angular momentum per unit mass:
\begin{equation}
    J = \frac{\mu v b}{\sqrt{1-v^2}},
\end{equation}
and with $E$ and $J$, we can define the impact parameter as
\begin{equation}
	b = \frac{J}{vE}.
\end{equation}
Using $ds^2=g_{\mu \nu}dx^{\mu}dx^{\nu} = -1$, which is the line element for the time-like particles, the orbit equation can be derived as
\begin{align} \label{eorb}
    F(u) \equiv \left(\frac{du}{d\phi}\right)^2 
    = \frac{C(u)^2u^4}{A(u)B(u)}\Bigg[\left(\frac{1}{vb}\right)^2-A(u)\left(\frac{1}{J^2}+\frac{1}{C(u)}\right)\Bigg],
\end{align}
which, in our case yields
\begin{equation} \label{eorb2}
    F(u)=\frac{1}{v^{2}b^2}+\left(\frac{1}{J^{2}}+u^{2}\right)\left(2mu-D(k,r_\text{c})\right).
\end{equation}
Here, $u = 1/r$ is usually done in celestial mechanics. Furthermore, note that we used Eq. \eqref{ee25} in this expression. Next, by an iterative method, the goal is to find $u$ as a function of $\varphi$, which we find as
\begin{equation} \label{euphi}
    u(\varphi) = \frac{\sin(\varphi)}{b}+\frac{1+v^2\cos^2(\varphi)}{b^2v^2}m.
\end{equation}
Note that we write $\varphi = \sqrt{D(k,r_\text{c})} \phi$. The Gaussian curvature can be derived using
\begin{align}
    K=-\frac{1}{\sqrt{g}}\left[\frac{\partial}{\partial r}\left(\frac{\sqrt{g}}{g_{rr}}\Gamma_{r\phi}^{\phi}\right)\right]
\end{align}
since $\Gamma_{rr}^{\phi} = 0$ for Eq. \eqref{eJac}. Furthermore, the determinant of Eq. \eqref{eJac} is
\begin{equation}
    g=\frac{B(r)C(r)}{A(r)^2}(E^2-\mu^2 A(r))^2.
\end{equation}
With the analytical solution to $r_\text{co}$, it is easy to see that
\begin{equation}
    \left[\int K\sqrt{g}dr\right]\bigg|_{r=r_\text{co}} = 0,
\end{equation}
which yields
\begin{equation} \label{gct}
    \int_{r_\text{co}}^{r(\varphi)} K\sqrt{g}dr = -\frac{A(r)\left(E^{2}-A(r)\right)C'-E^{2}C(r)A(r)'}{2A(r)\left(E^{2}-A(r)\right)\sqrt{B(r)C(r)}}\bigg|_{r = r(\varphi)},
\end{equation}
where the prime denotes differentiation with respect to $r$. The weak deflection angle is then \cite{Li:2020wvn},
\begin{align} \label{eqwda}
    \hat{\alpha} = \int^{\varphi_\text{R}}_{\varphi_\text{S}} \left[-\frac{A(r)\left(E^{2}-A(r)\right)C'-E^{2}C(r)A(r)'}{2A(r)\left(E^{2}-A(r)\right)\sqrt{B(r)C(r)}}\bigg|_{r = r(\varphi)}\right] d\phi + \phi_\text{RS}.
\end{align}
Using Eq. \eqref{euphi} in Eq. \eqref{gct}, we find
\begin{align} \label{gct2}
    \left[\int K\sqrt{g}dr\right]\bigg|_{r=r(\varphi)} &=-\phi_\text{RS}\sqrt{D(k,r_\text{c})} -\frac{\left(2E^{2}-1\right)m(\cos(\varphi_\text{R})-\cos(\varphi_\text{S}))}{\left(E^{2}-1\right)b \sqrt{D(k,r_\text{c})}}
\end{align}
We obtained the solution for $\phi$ as
\begin{align} \label{ephi}
    \varphi_\text{S} &=\arcsin(bu) + \frac{m\left[v^{2}\left(b^{2}u^{2}-1\right]-1\right)}{bv^{2}\sqrt{1-b^{2}u^{2}}}, \nonumber\\
    \varphi_\text{R} &=\pi-\arcsin(bu) - \frac{m\left[v^{2}\left(b^{2}u^{2}-1\right]-1\right)}{bv^{2}\sqrt{1-b^{2}u^{2}}}.
\end{align}
With the above expression for $\varphi$, we apply some basic trigonometric properties:
\begin{align}
    \cos(\pi-\varphi_\text{S})=-\cos(\varphi_\text{S}),
\end{align}
and we find the following:
\begin{align} \label{ecos}
    \cos(\varphi_\text{S}) &= \sqrt{1-b^{2}u^{2}}-\frac{m u\left[v^{2}\left(b^{2}u^{2}-1\right)-1\right]}{v^2\sqrt{\left(1-b^{2}u^{2}\right)}}.
\end{align}
Finally, using Eqs. \eqref{ecos} to Eq. \eqref{gct2} and noting Eq. \eqref{eqwda}, we derived the weak deflection angle for both time-like and null particles as
\begin{align} \label{ewda}
    \hat{\alpha} &= \Bigg\{\arcsin(bu_\text{S}) + \arcsin(b u_\text{R}) + \frac{m}{bv^2}\left[\frac{v^{2}\left(b^{2}u_\text{S}^{2}-1\right)-1}{\sqrt{1-b^{2}u_\text{S}^{2}}} + \frac{v^{2}\left(b^{2}u_\text{R}^{2}-1\right)-1}{\sqrt{1-b^{2}u_\text{R}^{2}}}\right] \Bigg\} \frac{1-D(k,r_\text{c})}{2} \nonumber \\
    &+ \frac{m\left(D(k,r_\text{c})v^{2}-D(k,r_\text{c})+2\right)}{b\sqrt{D(k,r_\text{c})}\left(D(k,r_\text{c})v^{2}-D(k,r_\text{c})+1\right)}\left(\sqrt{1-b^{2}u_\text{S}^{2}}+\sqrt{1-b^{2}u_\text{R}^{2}}\right)
\end{align}
which is general since it also admits a finite distance of the source and the receiver from the black hole. We see that there is no apparent divergence occurring due to the soliton DM contribution $D(k,r_\text{c})$. If $u_\text{R}$ and $u_\text{S}$ are very close to zero, we can recast the above equation into its far approximation form, which is
\begin{align} \label{ewda2}
    \hat{\alpha} &\sim \frac{2m\left(D(k,r_\text{c})v^{2}-D(k,r_\text{c})+2\right)}{b\sqrt{D(k,r_\text{c})}\left(D(k,r_\text{c})v^{2}-D(k,r_\text{c})+1\right)}+\frac{\left(1-\sqrt{D(k,r_\text{c})}\right)m\left(1+v^{2}\right)}{bv^{2}}.
\end{align}
For null case only, we have ($v = 1$),
\begin{align} \label{ewda3}
    \hat{\alpha} \sim \frac{4m}{b\sqrt{D(k,r_\text{c})}}-\frac{2m\left(1-\sqrt{D(k,r_\text{c})}\right)}{b}.
\end{align}
Clearly, if there is no soliton dark matter contribution ($k=0$), then $D(k,r_\text{c}) = 1$ and the above simply reduces to the Schwarzschild case.
\begin{figure*}
    \centering
    \includegraphics[width=0.48\textwidth]{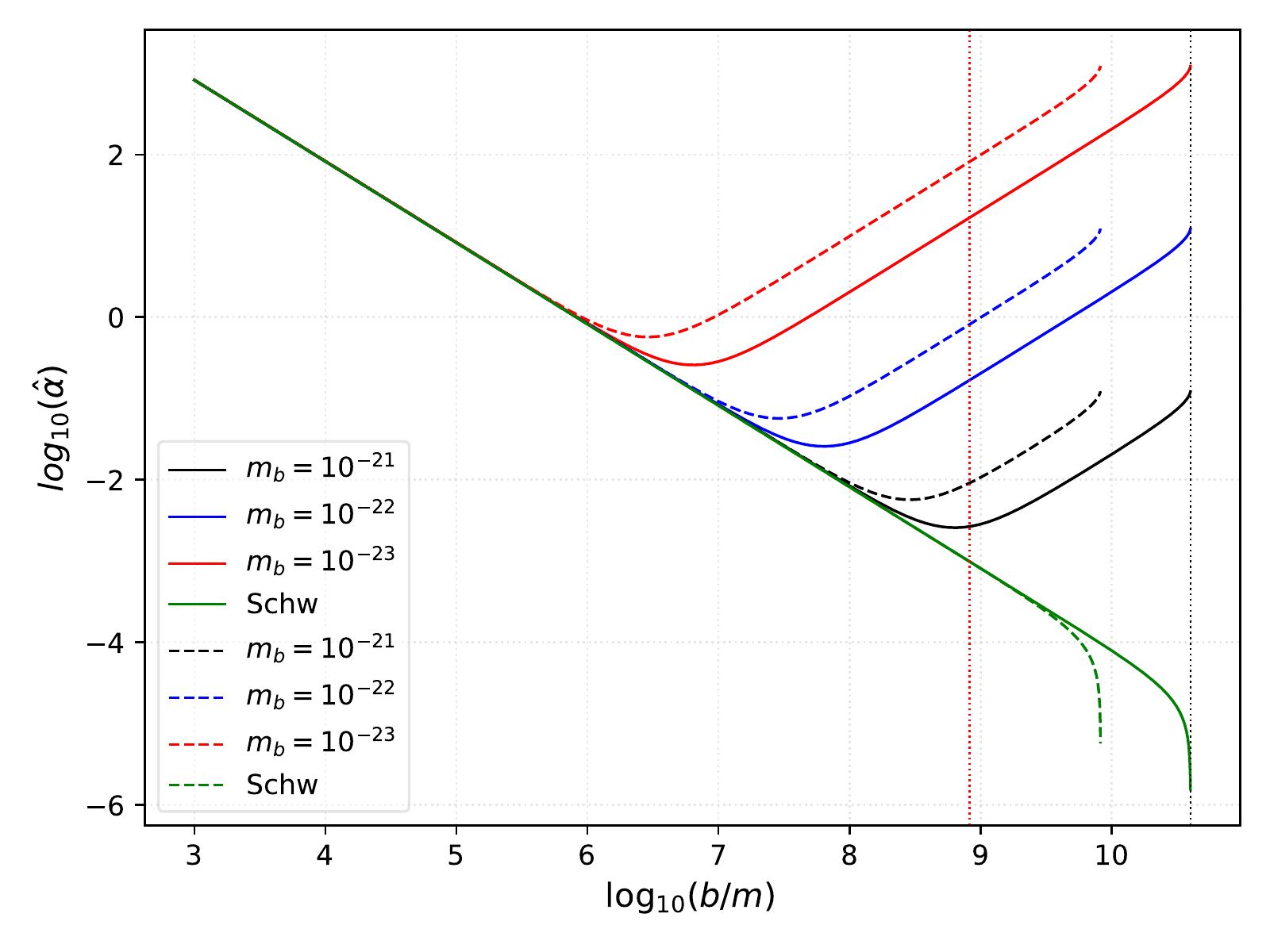}
    \includegraphics[width=0.48\textwidth]{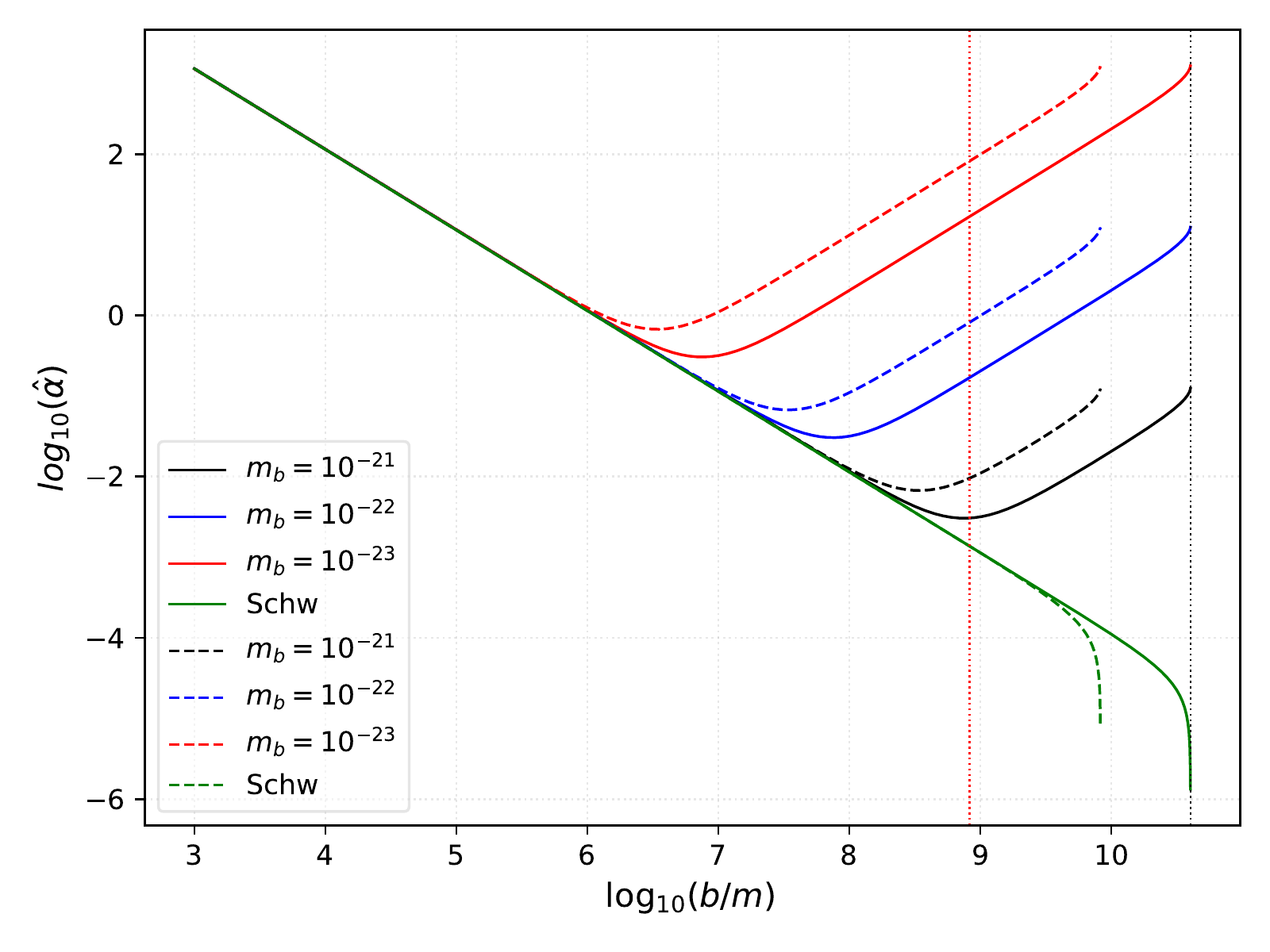}
    \caption{Behavior of weak deflection (in $\mu$as) under the influence of soliton profile. Here, we used the data for Sgr. A*, and assumed that $u_\text{S} = u_\text{R}$. In each plot, the solid line indicates $u_\text{R} = m/(8277\text{ pc})$ while the dashed line indicates $u_\text{R} = m/(1700\text{ pc})$. Lastly, the left plot is for $v = 1$, while the right plot is for $v = 0.75$. The red and black vertical dotted lines correspond to $r_\text{c}$ and the actual distance of Earth from Sgr. A*, respectively. }
    \label{fig_wda}
\end{figure*}
We plot the results in Fig. \ref{fig_wda}, where the general case in Eq. \eqref{eqwda} is applied. We only considered Sgr. A* in this case since the trend is fairly much the same for M87*. Overall, we see that the timelike particles (right panel) produce a slightly greater value for $\hat{\alpha}$ than the null case. Compared to the Schwarzschild case, the general effect of the boson mass is to increase $\hat{\alpha}$ with $m_\psi = 10^{-23}$ eV giving the highest value. We also see the effect of finite distance. For instance, if the receiver near the soliton core $r_\text{c}$ (represented by the dashed lines), $\hat{\alpha}$ is indeed greater as compared to the distant observer (solid lines). We could tell from these plots that at our location, the case of $m_\psi = 10^{-22}$ eV gives around $\hat{\alpha} = 12.19\mu$as provided that the impact parameter is comparable to $1/u_\text{R}$. Interestingly, this same value of $\hat{\alpha}$ occurs inside the soliton core when the impact parameter is $b/m \sim 67600$. In addition, as we decrease $b/m$, the $\hat{\alpha}$ decreases and a turning point occurs inside the soliton core. If $b/m << r_\text{c}/m$, we see that the behavior of $\hat{\alpha}$ follows the Schwarzschild trend and such an increase due to the soliton effect is very tiny. To conclude this discussion, our results suggest that the deviations due to the soliton dark matter can be noticeable near the soliton core.

\subsection{Einstein Rings}
One application of the weak deflection angle is the formation of the Einstein ring. Let us now calculate and form an estimate to find out the angular size of the Einstein rings due to the effect of the solitonic profile. First, let $D_\text{S}$ and $D_\text{R}$ be the distance of the source and the receiver, respectively from the lensing object which is the black hole. The thin lens condition states that $D_\text{RS}=D_\text{R}+D_\text{S}$. To find the position of the weak field images, we consider the lens equation given as \cite{Bozza:2008ev}
\begin{equation}
    D_\text{RS}\tan\beta=\frac{D_\text{R}\sin\theta-D_\text{S}\sin(\hat{\alpha}-\theta)}{\cos(\hat{\alpha}-\theta)}.
\end{equation}
If $\beta=0$, an Einstein ring is formed. The above equation gives the ring's angular radius as
\begin{equation}
    \theta_\text{Eins}\sim\frac{D_\text{S}}{D_\text{RS}}\hat{\alpha}.
\end{equation}
Furthermore, the relation $b=D_\text{R}\sin\theta \sim D_\text{R}\theta$ if the Einstein ring is assumed to be very small. We then find the angular size of the solitonic Einstein ring is
\begin{equation} \label{e47}
    \theta_\text{Eins}^{\text{sol}}=\sqrt{\frac{D_\text{S}}{D_\text{RS}D_\text{R}}\left[\frac{4m}{\sqrt{D(k,r_\text{c})}} - 2m\left(1-\sqrt{D(k,r_\text{c})}\right) \right]}.
\end{equation}
From the galactic center to our location, the distance is approximately $D_\text{R}\sim 8277$ pc, which justifies the use of the Eq \eqref{ewda3}. For M87*, this is $D_\text{R}\sim 16.8$ Mpc. Without the influence of dark matter, the angular radius depends solely on the source's distance from the lensing object if $D_\text{R}$ is fixed.
\begin{figure*}
    \centering
    \includegraphics[width=0.48\textwidth]{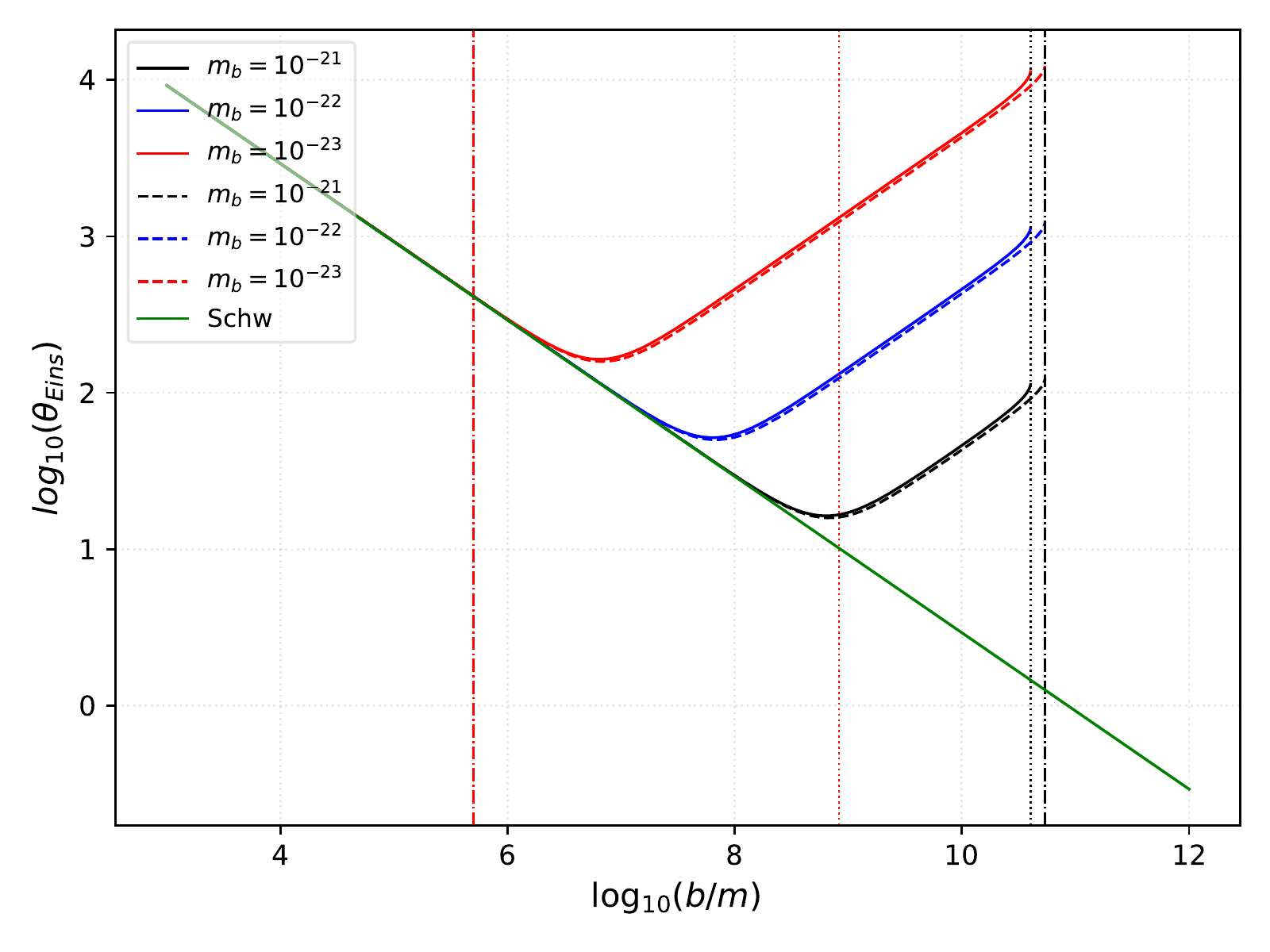}
    \caption{The angular radius of the Einstein ring (in $\mu$as) under the influence of soliton profile. Here, we used the data for Sgr. A* (solid lines) and M87* (dashed lines). We assumed in this plot that $D_\text{S} = D_\text{R}$, and the impact parameter is $\sim D_\text{R}$. The red vertical dotted and dash-dot lines correspond to the soliton cores of Sgr. A* and M87*, respectively. The black vertical dotted and dash-dot lines correspond to our location from Sgr. A* and M87*, respectively. }
    \label{einsring}
\end{figure*}
We visualize the result in Fig. \ref{einsring} to see how the Einstein ring behaves as the soliton core changes. Taking $m_\psi = 10^{-22}$ eV as an example, we saw earlier in Fig. \ref{fig_wda} that $\hat{\alpha} = 12.19\mu$as for Sgr. A* when $r_\text{c}/m = 170$ pc. The Einstein ring gives us a higher value, around $\theta_\text{Eins}^{\text{sol}} = 1135\mu$as. Take note that this value is for an impact parameter $\sim 8277$ pc. More interesting is the Einstein ring formation of objects which are close to BH and with low impact parameters. For this case, Fig. \ref{einsring} still gives such information. At low impact parameters, inside the soliton core, we can see the minimum of the curve due to the soliton dark matter effects, which deviates considerably from the Schwarzschild case. For example, taking $m_\psi = 10^{-22}$ eV, the minimum occurs near $b/m \sim 30$ pc in Sgr. A*, and $b/m \sim 0.015$ pc in M87*. The corresponding values for the Einstein rings are $\theta_\text{Eins}^{\text{sol}} = 51.4\mu$as and $\theta_\text{Eins}^{\text{sol}} =49.9\mu$as for Sgr. A* and M87*, respectively. Note that even when M87* is located at a vast distance compared to Sgr. A*, the difference is just small. Furthermore, these values are considerably higher compared to the Schwarzschild case where at such an estimated impact parameter, the value is $\theta_\text{Eins}^{\text{Schw}} \sim 36.3\mu$as. While these deviations occur inside the soliton core of Sgr. A*, it occurs outside the soliton core of M87*. Finally, these values are sufficient and more than enough to be detected by modern astronomical/space detectors such as the EHT, which can achieve an angular resolution of 10 − 15μas within 345 GHz in the future. Furthermore, the ESA GAIA mission is capable of resolving around $20\mu$as - $7\mu$as \cite{Liu:2016nwt}, and more powerful space-based technology called the VLBI RadioAstron in the future \cite{Kardashev:2013cla} can obtain a smaller angular resolution ranging from $1-10\mu$as. We remark that deeper in the soliton core, the difference between the deviation caused by the soliton dark matter relative to the Schwarzschild case becomes negligibly small. Thus, Einstein rings are better to be detected around the vicinity of the soliton core boundary.

\section{Conclusion} \label{conc}
In this paper, we derived a new metric that unifies the geometries of the central SMBH and the soliton (fuzzy) dark matter that surrounds it. In particular, we examined the effect of boson mass $m_\psi$ and the soliton $r_\text{c}$ as two free parameters to the SMBH in the Milky Way and M87 galaxies. The initial goal was to find constraints for $r_\text{c}$ given that $10^{-23}\text{ eV}\leq m_\psi \leq 10^{-21}\text{ eV}$. We have chosen this range since these are the values that are very close to the constraints found in the literature \cite{Chen:2016unw,Calabrese:2016hmp,Wasserman:2019ttq,Amorisco:2018dcn,Davies:2019wgi,Bar:2019bqz,Bar:2021kti,Corasaniti:2016epp,Irsic:2017yje,Leong:2018opi,Schive:2014dra,Li:2020qva} using different astrophysical observations and simulations. We found out that the effect of the soliton profile is to both increase the photonsphere and shadow radii. The minimum value of $r_\text{c}$ is also larger for smaller values of the boson mass. For constraint result, see Table \ref{tab2}. Next, we also examine the actual behavior of the shadow radius as the location of the static observer changes radially. We found out that just outside $r_\text{c}$, the shadow radius is seen to increase slightly as compared to the shadow radius inside, which is a piece of evidence that the soliton mass contained within the core radius acts as an effective mass. Such a fluctuation may be experimentally feasible for detection.

To gain more insights into the detectability of the soliton dark matter effects, we also considered analyzing the weak deflection angle and the Einstein ring it produced. We used the realistic parameters for Sgr. A* and M87*. We observe that for very large impact parameters, the weak deflection angle increases drastically in the Schwarzschild case. It merely implies that the soliton mass serves as an additional effective mass to the black hole, which might explain such a behavior. In the context of SMBH at the galactic centers, it is more interesting to use sources of light where the impact parameters are comparable to the soliton core. In this case, we found a slight deviation from the Schwarzschild case. Finally, we found that the angular radius of the Einstein ring is greater than the weak deflection angle, and Sgr. A* provides a slightly greater value than in M87*.

Research prospects include the following: $(1)$ Consider the spin parameter $a$ of the black hole and perform a detailed analysis. While it is acceptable to constraint $r_\text{c}$ in this study using $a=0$ (based on the arguments present in Refs. \cite{Vagnozzi:2022moj,EventHorizonTelescope:2021dqv}), it would be interesting to see changes in $r_\text{c}$ for the case where $a \neq 0$; $(2)$ The halo-core relation used in this study is the simplest one. One may consider, say, determining the halo-core relation under the influence of the nuclear bulge \cite{Li:2020qva,Davies:2019wgi} and see how will it affect the black hole geometry; $(3)$ Recently, there are other methods considered by various authors concerning the effect of the dark matter halo on a black hole at the galactic center \cite{Konoplya:2022hbl,Perivolaropoulos:2019vgl,Cardoso:2021wlq}. It would be interesting to perform a comparison between these methods, generating a black hole solution with the soliton profile, and analyze its black hole properties.

\begin{acknowledgements}
 A. {\"O}. and R. P. would like to acknowledge networking support by the COST Action CA18108 - Quantum gravity phenomenology in the multi-messenger approach (QG-MM).
\end{acknowledgements}

\bibliography{ref}
\bibliographystyle{apsrev}
\end{document}